\documentclass[conference]{IEEEtran}

\usepackage{xparse}

\usepackage{booktabs}
\usepackage{xcolor, soul}
\usepackage{xspace}
\usepackage{algorithm}
\usepackage[noend]{algpseudocode}
\usepackage{subfig}
\usepackage{multirow}
\usepackage{enumitem}
\usepackage{cite}
\usepackage{amsmath,amssymb,amsfonts}
\usepackage{graphicx}
\usepackage{textcomp}
\usepackage{url}
\usepackage[font=small,labelfont=bf]{caption}
\usepackage{amsfonts}
\usepackage{amssymb}
\usepackage{mathtools}
\DeclarePairedDelimiter{\ceil}{\lceil}{\rceil}
\newcommand{\mat}[1]{\mathbf{#1}}
\newcommand{\flop}{\mathrm{flops}}
\newcommand{\nrow}{\textit{nrows}}
\newcommand{\ncol}{\textit{ncols}}
\newcommand{\nnz}{\textit{nnz}}
\newcommand{\cf}{\textit{cf}}

\newcommand{\mA}{\mathbf{A}} 
\newcommand{\mB}{\mathbf{B}}

\newcommand{\transpose}     {^{\mbox{\scriptsize \sf T}}}

\newcommand{\mC}{\mathbf{C}}
\newcommand{\mD}{\mathbf{D}}

\newcommand{\revision}[1]{{\color{black} #1}}

\def\BibTeX{{\rm B\kern-.05em{\sc i\kern-.025em b}\kern-.08em
    T\kern-.1667em\lower.7ex\hbox{E}\kern-.125emX}}
\begin{document}

\title{\huge{Communication-Avoiding and Memory-Constrained\\ Sparse Matrix-Matrix Multiplication at Extreme Scale}}

\author{\IEEEauthorblockN{
    Md Taufique Hussain\IEEEauthorrefmark{1},
    Oguz Selvitopi\IEEEauthorrefmark{2},
    Aydin Bulu\c{c}\IEEEauthorrefmark{3}
    and
    Ariful Azad\IEEEauthorrefmark{4} }  
  \IEEEauthorblockA{\IEEEauthorrefmark{1}
    Indiana University, Bloomington, IN (mth@indiana.edu)}
  \IEEEauthorblockA{\IEEEauthorrefmark{2}
    Lawrence Berkeley National Laboratory, Berkeley, CA (roselvitopi@lbl.gov)}
  \IEEEauthorblockA{\IEEEauthorrefmark{3}
    Lawrence Berkeley National Laboratory, Berkeley, CA (abuluc@lbl.gov)}
  \IEEEauthorblockA{\IEEEauthorrefmark{4}
    Indiana University, Bloomington, IN (azad@iu.edu)}
}  
  
\renewcommand\citepunct{, }

\maketitle

%
\begin{abstract}
Sparse matrix-matrix multiplication (SpGEMM) is a widely used kernel in various graph, scientific computing and machine learning algorithms.  
In this paper, we consider SpGEMMs performed on hundreds of thousands of processors generating trillions of nonzeros in the output matrix. 
Distributed SpGEMM at this extreme scale faces two key challenges: (1) high communication cost and (2) inadequate memory to generate the output. 
We address these challenges with an integrated communication-avoiding and memory-constrained SpGEMM algorithm that scales to 262,144 cores (more than 1 million hardware threads) and can multiply sparse matrices of any size as long as inputs and a fraction of output fit in the aggregated memory. 
As we go from 16,384 cores to 262,144 cores on a Cray XC40 supercomputer, the new SpGEMM algorithm runs 10x faster when multiplying large-scale protein-similarity matrices.
\end{abstract}


%
\maketitle

\section{Introduction}
Multiplication of two sparse matrices (SpGEMM) is a common operation in numerous computing fields including data analytics~\cite{he2010parallel, jamour2019matrix}, graph processing~\cite{Azad2015, solomonik2017scaling, combblas2011}, bioinformatics~\cite{jain_et_alWABI19, guidi2018bella}, machine learning~\cite{qin2020sigma},
computational chemistry~\cite{borvstnik2014sparse}, and scientific computing~\cite{mccourt2015sparse}. 
Most use cases of SpGEMM have low arithmetic intensity, resulting in poor scaling to large concurrencies. 
Despite these inherent challenges, the need to scale SpGEMM on large distributed-memory clusters for solving important science 
problems generated a fruitful line of research.
As a result, the last decade saw the development of increasingly sophisticated distributed-memory parallel SpGEMM algorithms that employ communication-avoiding techniques, perform communication overlapping, and partitioning the matrices to minimize communication~\cite{Akbudak2018, ballard2016hypergraph}.

Different from other matrix multiplication instances such as dense-dense and sparse-dense, SpGEMM has four unique features that make distributed SpGEMM harder to develop.
(1) The output of SpGEMM often has more nonzeros than the inputs combined. In some applications, the output of SpGEMM may not even fit in the aggregate memory of large supercomputers. 
(2) the number of nonzeros of the output matrix is not known a-priori for SpGEMM. Hence, a distributed symbolic step is needed to estimate the memory required for the multiplication.  
(3) SpGEMM typically has low arithmetic intensity. As a result, local computations on each node becomes computationally expensive. 
(4) At extreme scale, inter-process communication becomes the performance bottleneck. 
Since the last two problems arise in all distributed SpGEMM, they are well studied in the literature~\cite{Akbudak2018, ballard2016hypergraph, Azad2016}. 
By contrast, the first two problems are relatively new to the community as SpGEMM performed with emerging biological and social networks has started to overrun the memory limit of supercomputers. 

This paper presents a distributed-memory algorithm and its high-performance implementation that harmoniously address all four challenges. 
(1) When the memory requirement of SpGEMM exceeds the aggregate memory, we multiply in $b$ batches, where a batch computes a subset of columns of the output.
(2) The required number of batches depends on the aggregate memory, input matrices, and the process grid used to distribute the matrix. 
We developed a distributed symbolic algorithm that efficiently identifies the memory requirements and the optimal number of batches needed.
(3) We integrate the batching technique within an existing  communication-avoiding framework~\cite{Azad2016} that multiplies matrices on a 3D process grid.
(4) We observe that in-node SpGEMM and merging routines needed by distributed SpGEMM do not need nonzeros sorted within columns of input and output matrices. In light of this observation, we developed new sort-free SpGEMM and merging algorithms to make local computations significantly faster than current state-of-the-art approaches.
Thus, this paper presents novel solutions to four challenges in large-scale SpGEMM, and it does so within the established communication-avoiding algorithms framework.

The proposed memory-constrained SpGEMM algorithm can be directly used with applications that do not access the whole output matrix all at once. 
For example, consider finding the Jaccard similarity between two datasets, a problem that is successfully formulated as multiplication of a sparse matrix with its transpose~\cite{besta2019communication}. BELLA sequence overlapper~\cite{guidi2018bella} and PASTIS many-to-many protein sequence aligner~\cite{sc20pastis} use the SpGEMM operation in a similar way. Since the subsequent analysis only needs to access subsets of the output $\mA \mA\transpose$, we can form it in batches, and discard each batch. In hypergraph partitioning using the effective multi-level paradigm, successively coarser graphs are generated. Prior to coarsening, one typically finds the number of shared hyperedges between all pairs of vertices in order to run a matching algorithm that identifies pairs of vertices to coarsen. This process is called heavy-connectivity matching~\cite{catalyurek1999hypergraph} or inner-product matching~\cite{karypis1999multilevel}, and can be formulated as another SpGEMM of type $\mA \mA\transpose$. Due to memory limitations and the higher density of the product, this SpGEMM is done in batches in distributed-memory multi-level partitioners such as Zoltan~\cite{devine2006parallel}. Finally, HipMCL~\cite{Azad2018} is a distributed-memory scalable version of the Markov clustering algorithm. HipMCL iterations involve matrix squaring followed by various pruning steps to reduce the memory footprint and increase convergence. In both HipMCL and hypergraph coarsening, columns of the output can be formed in batches, and immediately used for pruning or matching without requiring the whole output.


In these scenarios, the higher-level application can form subsets of the output matrix in batches, perform the required computation on it, and discard some or all of its nonzeros before moving into the next batch. The solutions provided in the literature for all these examples are ad-hoc algorithms. The community lacks a thorough understanding of theoretical and practical performance that can be expected from a memory-constrained SpGEMM that operates in batches. This results in late discoveries of performance and scalability problems in large-scale runs. Our work fills that crucial gap.


Our integrated communication-avoiding (CA) and memory-constrained SpGEMM algorithm is generally applicable even when forming the full output is feasible. In that case, our algorithm will choose to form the output all at once if that minimizes communication. Our algorithm strong scales naturally well because the increase in the available memory allow our integrated algorithm to either decrease the number of passes over the input, or increase replication in exchange of reduced communication, or do both. 

Overall, the faster local computation, reduced communication and batching techniques delivered a massively parallel algorithm that scales to millions of threads on a Cray XC40 supercomputer.
Using batching, our SpGEMM can multiply massive matrices using 0.5 PB memory whereas previous SpGEMMs would have required 2.2 PB memory, thereby could not solve the problem at all.


\revision{
The main contributions of the paper are summarized below.
\begin{enumerate}[itemsep=0pt,topsep=4pt, leftmargin=4mm]

    \item We develop a new distributed-memory SpGEMM algorithm that multiplies matrices in batches when the required memory to generate the output exceeds the available memory. 
    This algorithm makes it possible to multiply sparse matrices of any size as long as inputs and a fraction of output fit in the memory.
    \item We adapt existing CA techniques in a memory-constrained setting. We develop new algorithms to reduce the overhead of CA algorithms due to batching.
    \item 
    We develop lower and upper bounds on the number of batches and present a symbolic algorithm to compute the exact number of batches needed. 
    
    \item We  demonstrate  that our  SpGEMM  algorithm  scales  to  4096  nodes  (more  than  1 million hardware threads) on a Cray XC40 supercomputer. The utility of the SpGEMM algorithm is demonstrated with large-scale protein similarity networks where the multiplication requires up to 300 trillion floating point operations and 2.2 PB memory.
\end{enumerate}
}

\section{Background and Notations}
\subsection{Notations}
\vspace{-2pt}
Given two sparse matrices $\mA {\in} \mathbb{R}^{m \times k}$ and ${\mB {\in} \mathbb{R}^{k \times n}}$, SpGEMM multiplies $\mA$ and $\mB$ and computes another potentially sparse matrix ${\mC {\in} \mathbb{R}^{m \times n}}$.
The operation can be also performed on an arbitrary semiring $\mathbb{S}$ instead of the field of real numbers $\mathbb{R}$ and our algorithms are applicable to that case as well since we do not utilize Strassen-like algorithms.  
In our analysis, we consider $n$-by-$n$ matrices (that is, $m{=}k{=}n$) for simplicity.
Given a matrix $\mA$, $\nrow(\mA)$ and $\ncol(\mA)$ denote the number of rows and columns in $\mA$.
$\nnz(\mA)$ denotes the number of nonzeros in $\mA$.
$\flop$ denotes the number of multiplications needed to compute $\mA\mB$.
The \emph{compression factor} ($\cf$) is the ratio of $\flop$ to $\nnz(\mC)$: $\cf {=} \flop/\nnz(\mC)$. 
Since at least one multiplication is needed for an output nonzero, $\cf{\geq} 1$.

\subsection{Problem Description}

In many data analytics applications that use SpGEMM,  the output matrix $\mC$ is often too large to store all at once in aggregate memory. 
Fortunately, the subsequent analysis rarely requires access to the whole $\mC$, so it can be formed in {\em batches}. Our paper is about this {\em memory-constrained SpGEMM} formulation where the available aggregate memory $M = \nnz(\mC)/k$ where $k>1$. We also assume that the inputs fit into aggregate memory, meaning $M > \nnz(\mA) + \nnz(\mB)$. Together, these two conditions imply $\nnz(\mC) > k(\nnz(\mA) + \nnz(\mB)) $ for $k>1$. 
Note that the actual memory requirement of distributed SpGEMM can be much larger than $O(\nnz(\mC))$ because of the need to store unmerged results in each process.
Our algorithm also covers the case where the final output fits in the memory, but the intermediate results exceed the available memory.

\subsection{Related work}
The algorithms developed for parallel SpGEMM stem from how the matrices and the
relevant computations are distributed across parallel computing units.
Most of the shared-memory parallel SpGEMM algorithms rely on Gustavson's
algorithm~\cite{Gustavson1978}, which yields a high level of parallelism as the
rows or the columns of the output matrix can be computed independent of each
other.
In the column variant, a number of columns of $\mA$ need to be scaled and accumulated in
order to compute the output column $\mC(:,j)$, i.e.,
\[
\mC(:,j) = \sum_{i:\mB(i,j) \neq 0} \mA(:,i) \mB(i,j).
\]
The algorithms usually differ in the data structure they use for accumulating
values.
Among them are the sparse accumulators~\cite{Gilbert1992, Patwary2015},
heaps~\cite{Liu2014, Azad2016}, hash tables~\cite{Deveci2017,Nagasaka2019}, and
merge sorts~\cite{Gremse2018, gu2020bandwidth}.
The accumulator choice has important implications on the performance of the
parallel algorithm and often depends on sparsity pattern
of the multiplied matrices, compression factor, and parallel architecture.


The studies regarding parallelization of SpGEMM on many- and multi-core systems
often focus on optimization aspects related to accumulators, efficient data
access, and load balancing.
As the many-core architectures necessitate a fine-grain load balancing for good
performance, several works on GPUs~\cite{Liu2014, Deveci2017, Nagasaka2017,
  Gremse2018} aim to achieve that goal.
The works for multi-core architectures~\cite{Patwary2015,Nagasaka2019} often
strive for improving the poor cache behavior with various techniques such as
blocking or using appropriate accumulators according to $\cf$.

The distributed memory algorithms for parallel SpGEMM can be categorized according to the data distribution method they use.
The algorithms that rely on one-dimensional (1D) distribution partition all matrices across the entire row or column dimension.
Although 1D distribution~\cite{Nusbaum2011} suffers from poor communication scalability, the communication costs can be reduced by pre-processing with graph/hypergraph partitioning models~\cite{Akbudak2018} that exploit the sparsity pattern of the multiplied matrices in order to reduce communicated data.
This pre-processing, however, can be prohibitive if SpGEMM is not utilized in a repeated context as the pre-processing stage often does not scale well.
CombBLAS~\cite{combblas2011} uses Sparse SUMMA algorithm~\cite{Buluc2012, Geijn1995} tailored for 2D distribution of matrices.
In 2D distribution, the matrices are partitioned into rectangular blocks and a 2D process grid is associated with the distribution.
The 2D distribution has better communication characteristics compared to 1D distribution.
In the 3D variant of the Sparse SUMMA algorithm, each sub-matrix is further divided  into layers.
This approach exhibits better scalability at larger node counts~\cite{Azad2016, lazzaro2017increasing}, where the multiplied instances become more likely to be latency-bound.
We examine these algorithms in detail in Sections~\ref{sec:summa2d}~and~\ref{sec:summa3d}, respectively, as they form the basis of our work.
Cannon's algorithm~\cite{cannon69thesis}, which uses a 2D distribution of matrices, has also been used in parallel SpGEMM~\cite{borvstnik2014sparse}.





\section{2D and 3D sparse SUMMA algorithms}
Our algorithmic framework is based on the 2D SUMMA algorithm~\cite{Buluc2012, Geijn1995}, and our communication-avoiding technique relies on the 3D SUMMA algorithm developed in prior work~\cite{Azad2016}.
In this section, we revisit 2D and 3D sparse SUMMA algorithms by putting them in a common framwork upon which the new memory-constrained algorithm will be built in the next section.


\begin{algorithm}[t]
\caption{An overview of 2D sparse SUMMA algorithm.}
\textbf{Input and Output:} Input matrices $\mat{A}$ and $\mat{B}$ and output matrix $\mat{C}$ are distributed on a 2D process grid $P_{2D}$. $\mat{\Tilde{A}}$, $\mat{\Tilde{B}}$, and $\mat{\Tilde{C}}$ denote local submatrices in the process considered.
\begin{algorithmic}[1]

\Procedure{Summa2D}{$\mat{A}, \mat{B}, P_{2D}$}
\For{all processes in $P_{2D}(i,j)$ {\bf in parallel}} 
    \State stages $\gets $ number of process columns in $P_{2D}$
    \For{ $s \gets 1$ to stages} \Comment{SUMMA stages}
     \State $\mat{\Tilde{A}}_{\text{recv}}$  $\gets$ \Call{Bcast}{$\mat{\Tilde{A}}, P_{2D}(i, s), P_{2D}(i, :)$}
     \State $\mat{\Tilde{B}}_{\text{recv}}$  $\gets$ $\Call{Bcast}{\mat{\Tilde{B}}, P_{2D}(s, j), P_{2D}(:, j)}$
     \vspace{2pt}
        
    \State $\mat{\Tilde{C}}[s]$  $\gets$ \Call{LocalMultiply}{$\mat{\Tilde{A}}_{\text{recv}}$, $\mat{\Tilde{B}}_{\text{recv}}$}
\EndFor
\State $\mat{\Tilde{C}}$ $\gets$ \Call{Merge}{$\mat{\Tilde{C}}[1..\text{stages}]$}
\EndFor
\State \Return $\mat{C}$
\EndProcedure
\end{algorithmic}
\label{algo:2Dsumma}
\end{algorithm}

\subsection{2D Sparse SUMMA}
\label{sec:summa2d}
{\bf Data distribution.} The original sparse SUMMA algorithm~\cite{Buluc2012} works on a 2D $p_r{\times}p_c$ process grid $P_{2D}$.  
In this paper, we only consider square process grid with $p_r{=}p_c$.  
$P_{2D}(i,j)$ denotes the process in $i$th row and $j$th column in the 2D grid.
$P_{2D}(i,:)$ denotes all processes in the $i$th row and $P_{2D}(:,j)$ denotes all processes in the $j$th column.

{\bf The algorithm.} The \textproc{Summa2D} function in Algorithm~\ref{algo:2Dsumma} describes the 2D multiplication algorithm that operates on matrices distributed on $P_{2D}$.
\textproc{Summa2D} proceeds in $p_c$ stages.
At stage $s$, each process $P_{2D}(i,j)$ participates in two broadcast operations in its process row and process column.
In the $i$th process row $P_{2D}(i,:)$, the process $P_{2D}(i,s)$ broadcasts its local submatrix $\Tilde{\mA}$ to all processes in $P_{2D}(i,:)$ (line 5, Alg.~\ref{algo:2Dsumma}).
Similarly, in the $j$th process column $P_{2D}(:,j)$, the process $P_{2D}(s,j)$ broadcasts its local submatrix $\Tilde{\mB}$ to all processes in $P_{2D}(;,j)$ (line 6, Alg.~\ref{algo:2Dsumma}).
Each process stores the received data in $\Tilde{\mA}_{\text{recv}}$ and $\Tilde{\mB}_{\text{recv}}$. 
In line 7 of Alg.~\ref{algo:2Dsumma}, we locally multiply the received matrices and generate a low-rank version of the output.
To facilitate merging at the end of all SUMMA stages, partial result from each stage is stored in an array. 
For example, at stage $s$, each process multiplies received input submatrices $\Tilde{\mA}_{\text{recv}}$ and $\Tilde{\mB}_{\text{recv}}$ and stores the result at $\Tilde{\mC}[s]$.
Finally, at the end of all SUMMA stages, each process merges $p_c$ pieces of partial results and creates its local piece $\Tilde{\mC}$ of the result (line 8, Alg.~\ref{algo:2Dsumma}). 
When \textproc{Summa2D} returns, all local pieces $\Tilde{\mC}$ form the output matrix $\mC$ distributed on $P_{2D}$.

{\bf Major steps.}
Each stage of \textproc{Summa2D} has three major steps: (a) {\em A-Broadcast}: broadcasting parts of $\mA$ along the process row;  (b) {\em B-Broadcast}: broadcasting parts of $\mB$ along the process column; and (c) {\em Local-Multiply}: performing multithreaded local multiplication. 
After all SUMMA stages, we perform another step {\em Merge} (will be called {\em Layer-Merge} in the 3D algorithm) that merges partial results. Here, merging means adding multiplied values with the same row and column indices.
While one can incrementally merge partial results after local multiplications, it is computationally more expensive in the worst case~\cite{selvitopi2020optimizing}. 
Hence, in this paper, we consider merging after completing all stages as shown in line 8 of  Alg.~\ref{algo:2Dsumma}.

\textproc{Summa2D} performs reasonably well on a few hundred processes. 
However, as the number of processes increases, communication (broadcasting $\mA$ and $\mB$) becomes the performance bottleneck of \textproc{Summa2D}~\cite{Buluc2012}. 
3D Sparse SUMMA~\cite{Azad2016} was developed to reduce the communication cost of SpGEMM.

\begin{algorithm}[t]
\caption{3D sparse SUMMA.}\label{euclid}
\textbf{Input and Output:} Input matrices $\mat{A}$ and $\mat{B}$ and output matrix $\mat{C}$ are distributed in a 3D process grid $P_{3D}$. $\mat{\Tilde{A}}$, $\mat{\Tilde{B}}$, and $\mat{\Tilde{C}}$ denote local submatrices in the current process. $\mat{A}^{(k)}$ denotes the 2D submatrix in the $k$th layer of $P_{3D}$.
\begin{algorithmic}[1]


\Procedure{Summa3D}{$\mat{A}, \mat{B}, P_{3D}$}

\For{all processes in $P_{3D}(i,j, k)$ {\bf in parallel}}
    \State $\mat{D}^{(k)} \gets$ \Call{Summa2D}{$\mat{A}^{(k)}$,  $\mat{B}^{(k)}$, $P_{3D}(:, :, k)$} 
    \vspace{1pt}
    \State $\mat{\Tilde{D}}^{(k)}[1..l]$ $\gets$ ColSplit($\mat{\Tilde{D}}^{(k)}, l$)
    \vspace{1pt}
    \State $\mat{\Tilde{C}}^{(k)}[1..l]$ $\gets$ AllToAll $(\mat{\Tilde{D}}^{(k)}[1..l]$, $P_{3D}(i, j, :))$
    \vspace{1pt}
    \State $\mat{\Tilde{C}}^{(k)}$ $\gets$ Merge($\mat{\Tilde{C}}^{(k)}[1..l])$
\EndFor
\State \Return $\mat{C}$
\EndProcedure
\end{algorithmic}
\label{algo:3Dsumma}
\end{algorithm}

\subsection{3D Sparse SUMMA}
\label{sec:summa3d}
{\bf Processes grid.} 
Here, each matrix is distributed in a 3D $p_r{\times}p_c{\times}l$ process grid $P_{3D}$.
All processes $P_{3D}(:,:,k)$ with a fixed value $k$ in the third dimension form the $k$th \emph{layer} of $P_{3D}$, where $1{\leq}k {\leq} l$. 
In this context, $P_{3D}(:,:,k)$ is similar to a 2D process grid discussed in Sec.~\ref{sec:summa2d}.
In this paper, we only consider square process grid in each layer.
Hence, with $p$ processes divided into $l$ layers, the shape of $P_{3D}$ is $\sqrt{p/l}\times \sqrt{p/l}\times l$.
Here, $P_{3D}(i,j, k)$ denotes the process in the $i$th row and $j$th column in the $k$th layer in $P_{3D}$.
All processes in $P_{3D}(i,j,:)$ form a \emph{fiber} that consists of  processes with the same row and column indices from different layers.

\begin{figure*}[!t]
    \centering
    \includegraphics[width=.85\linewidth]{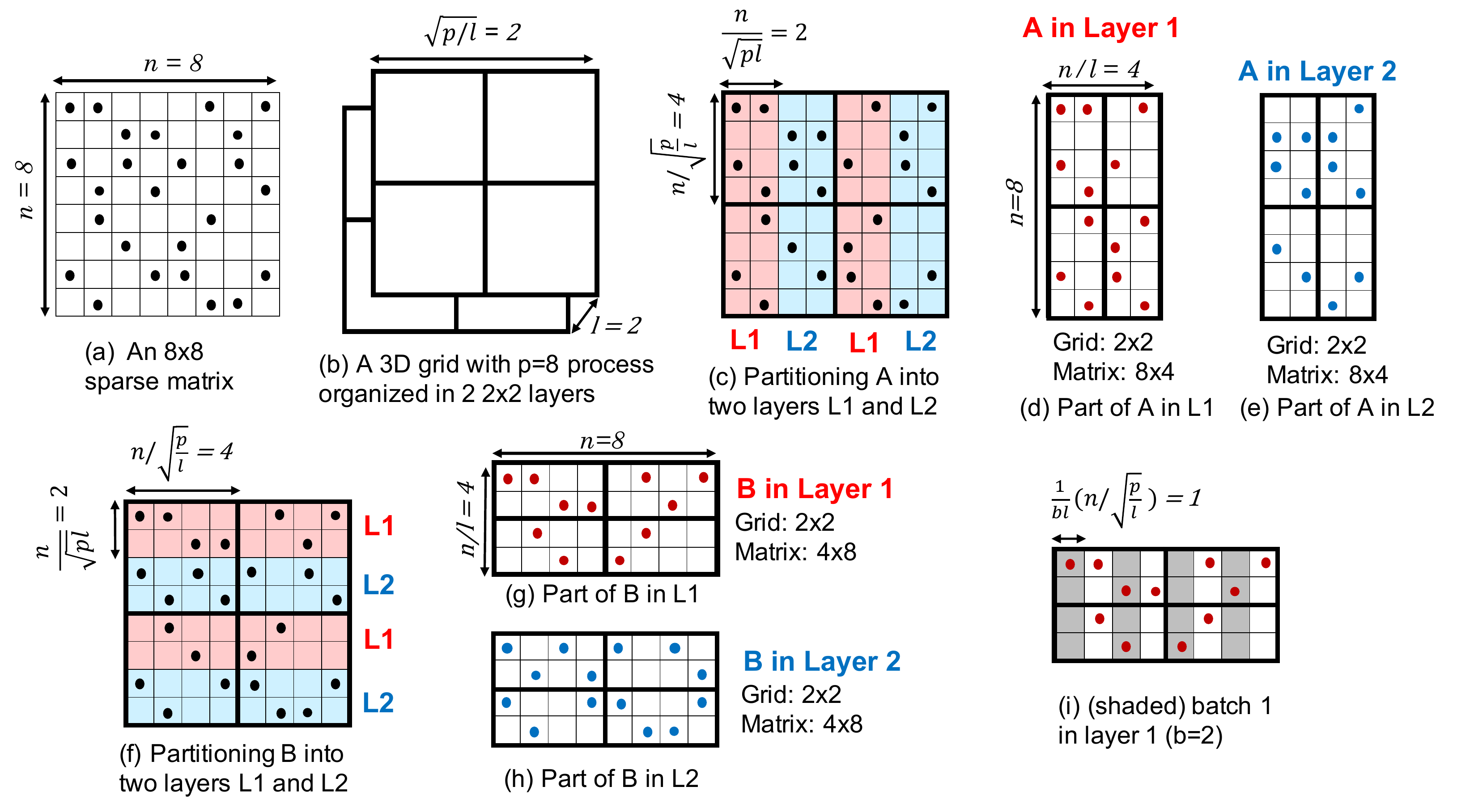}
    \vspace{-5pt}
    \caption{
    Distributing an input matrix shown in (a) into a $2\times 2\times 2$ 3D process grid shown in (b). Here, we use the same matrix for both $\mA$ and $\mB$. Thick borders denotes process boundaries in each $2\times 2$ layer. (c) $\mA$ is partitioned along the column to create layers. Red slices form layer 1 and blue slices form layer 2. (d, e) Rectangular submatrices of $\mA$ in layer 1 and 2. (f) $\mB$ is partitioned along the row to create layers. Red slices form layer 1 and blue slices form layer 2. (g, h) Rectangular submatrices of $\mB$ in layer 1 and 2. (i) Assuming $b=2$, the first batch in layer 1 of $\mB$ is shown by shaded regions.}
     \vspace{-5pt}
    \label{fig:data_distribution}
\end{figure*}

{\bf Data distribution.} 
Suppose $\mA$ is an $n\times n$ matrix. After distributing $\mA$ in a 3D grid, let  $\mA^{(k)} \in \mathbb{R}^{n \times (n/l)}$ be the submatrix of $\mA$ distributed in the $k$th layer.
Fig.~\ref{fig:data_distribution}(c) shows that each layer gets slices of $\mA$ that respect the 2D process boundary. 
After 3D distribution, each local piece $\Tilde{\mA}$ is an  $\bigl(n/\sqrt{p/l}\bigr) \times \bigl(n/\sqrt{pl}\bigr)$ submatrix. 
Hence, $\nrow(\Tilde{\mA}) = l \cdot \ncol(\Tilde{\mA})$ for all local submatrices $\Tilde{\mA}$.
Similarly, we split $\mB$ along the rows, and each local piece $\Tilde{\mB}$ is an  $\bigl(n/\sqrt{pl}\bigr) \times \bigl(n/\sqrt{p/l}\bigr)$ submatrix (Fig.~\ref{fig:data_distribution}(f)).
As $l$ increases, $\Tilde{\mA}$ becomes tall and skinny and $\Tilde{\mB}$ becomes short and fat. 
Fig.~\ref{fig:data_distribution} shows an example with a 3D grid where $p=8$ processes are organized in $l=2$ layers, and each layer is a $2{\times}2$ grid.
Since $\mA$ and $\mB$ are distributed differently in $P_{3D}$, we chose to distribute $\mC$ similar to $\mA$.

\begin{table}[!t]
    \centering
    \caption{Symbols used in this paper}
    \begin{tabular}{ll}
    \toprule
    Symbol & Meaning \\
    \toprule
         $p$  & total number of processes \\
         $l$ & number of layers in a 3D grid \\ 
         $b$ & number of batches \\ 
         $P_{3D}$ & a $\sqrt{p/l}\times \sqrt{p/l}\times l$ process grid \\
        
         $\mA$ & first input matrix (distributed) \\
         $\mB$ & second input matrix (distributed) \\
         $\mC$ & output matrix (distributed) \\
         $\mA^{(k)}$ & Part of $\mA$ in the $k$th layer (similarly $\mB^{(k)}$ and $\mC^{(k)}$)\\
         $\mD^{(k)}$ & A low-rank version of output matrix in the $k$th layer \\
         $\Tilde{\mA}$ & parts of $\mA$ stored in the current process (similarly $\Tilde{\mB}$, $\Tilde{\mC}$) \\
         $n$ & total number of rows/columns in $\mA, \mB, \mC$ \\ 
         $M$ & total available memory in all processes \\
         $r$ & number of bytes needed to store a nonzero (on average)\\
         
    \bottomrule
    \end{tabular}
    \vspace{-10pt}
    \label{tab:my_label}
\end{table}

\begin{figure*}[!t]
    \centering
    \includegraphics[width=.78\linewidth]{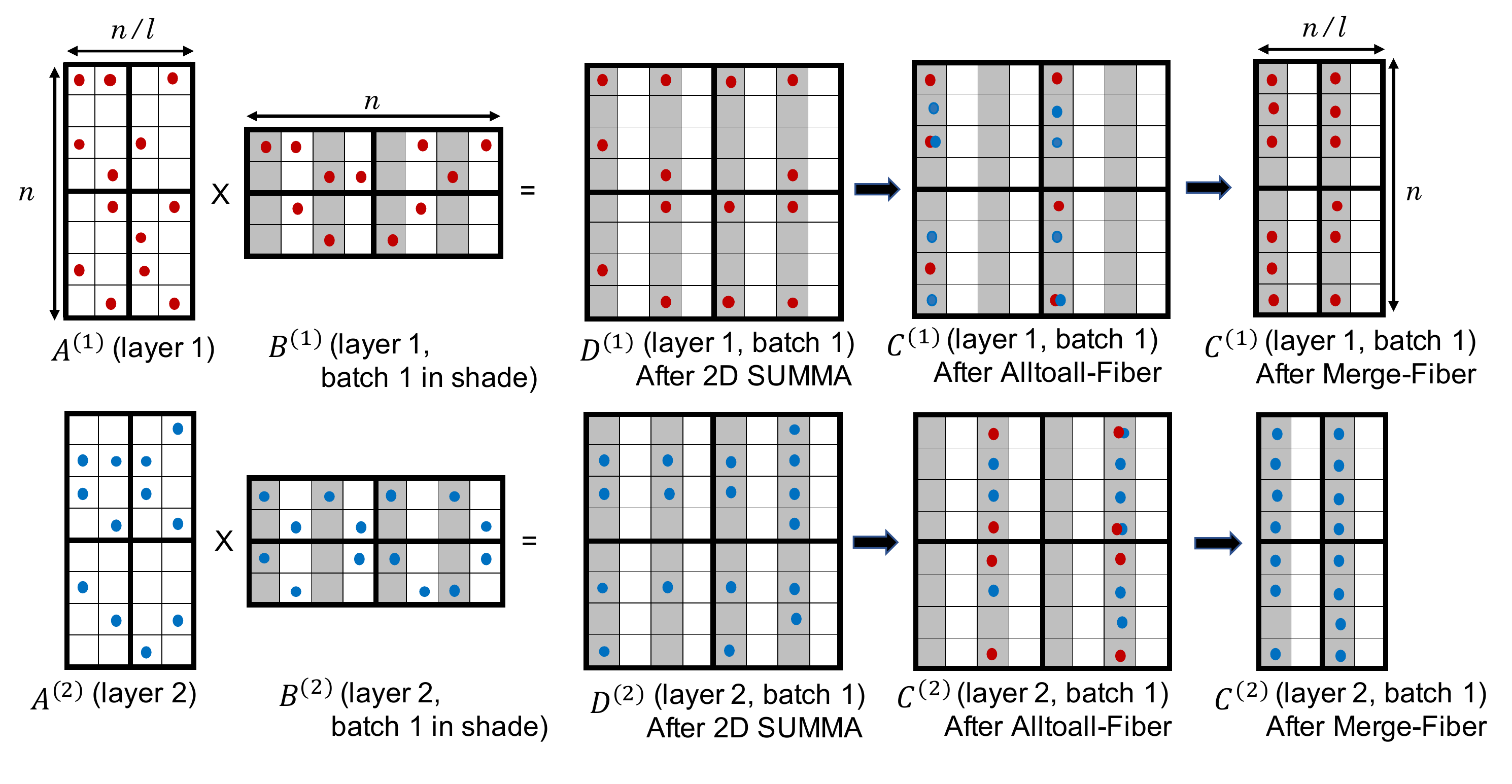}
    \vspace{-5pt}
    \caption{Assuming $b=2$, major steps of the first batch of Alg.~\ref{algo:batched3Dsumma} are shown when multiplying the input matrices shown in Fig.~\ref{fig:data_distribution}. Columns involved in this batch are shown by shaded regions. Red dots represent nonzeros in layer 1 and blue dots represent nonzeros in layer 2. 
    After performing 2D multiplication in each layer, we obtain $\mD^{(1)}$ and $\mD^{(2)}$ at layer 1 and 2.    After AllToall-Fiber both layers have some red and some blue dots as each process exchanges along the fiber some part of its result obtained from \textproc{Summa2D}. Overlapping red and blue dots indicate that result of 2D SUMMA in each layer had nonzeros at those positions. These overlapped entries are merged during Merge-Fiber step to have final result of the first batch. Similarly, we can compute the result of the second batch using unshaded parts of $\mB$.}
    \vspace{-5pt}
    \label{fig:BatchedSUMMA3D}
\end{figure*}

{\bf The algorithm.}
The \textproc{Summa3D} function in Algorithm~\ref{algo:3Dsumma} describes the 3D multiplication algorithm that operates on matrices distributed on $P_{3D}$. Initially, \textproc{Summa3D} proceeds independently at each layer $P_{3D}(:,:,k)$ by calling \textproc{Summa2D} with 2D input  matrices in that layer (line 3, Alg.~\ref{algo:3Dsumma}). 
At each layer, \textproc{Summa2D} multiplies $\mA^{(k)}\in \mathbb{R}^{n \times \frac{n}{l}}$ and $\mB^{(k)}\in \mathbb{R}^{\frac{n}{l} \times n}$ to produce $\mD^{(k)} \in \mathbb{R}^{n\times n}$.
Here, $\mD^{(k)}$ denotes an intermediate low-rank output matrix that needs to be merged across layers to form the final product. 
Next, each process splits $\mat{\Tilde{D}}^{(k)}$, the local submatrix of $\mD^{(k)}$, into $l$ parts $\mat{\Tilde{D}}^{(k)}[1..l]$ by splitting $\mat{\Tilde{D}}^{(k)}$ along the column (line 4, Alg.~\ref{algo:3Dsumma}). Then, each process performs an AllToAll operation on every fiber $P_{3D}(i,j,:)$ (line 5, Alg.~\ref{algo:3Dsumma}). Finally, each process merges all the copies received from the AllToall operation in previous step to form the final local copy of the result matrix (line 6, Alg.~\ref{algo:3Dsumma}).
The last two steps are called AllToAll-Fiber and Merge-Fiber.


\section{New algorithms for \\memory-constrained 3D SpGEMM}

\subsection{The case for batching}
{\bf Memory requirement of 3D Sparse SUMMA.}
If $r$ bytes are needed to store a nonzero, the memory requirement to perform SpGEMM is a least $r(\nnz(\mA) + \nnz(\mB) + \nnz(\mC))$ bytes.
However, a distributed algorithm requires much more memory in practice. 
The \textproc{Summa3D} function from Algorithm~\ref{algo:3Dsumma} on a $\sqrt{p/l}\times \sqrt{p/l}\times l$ process grid stores unmerged matrices $\mD^{(k)}$ at layer $k$.
Thus, we have the following inequality:
\vspace{-3pt}
\begin{equation}
    \flop \geq \sum_{k=1}^l\nnz(\mD^{(k)}) \geq \nnz(\mC).
    \label{eq:mem3d}
\end{equation}
Here, the Merge-Layer operation will generate $\mD^{(k)}$ at every layer, and the Merge-Fiber will produce $\mC$.
In the worst case, we may need to store $\flop$ nonzeros when no merging happens inside Local-Multiply in \textproc{Summa2D}.
Let mem($\mC$) denote the memory requirement for the \textproc{Summa3D} function. 
Then, $\text{mem}(\mC)= r \sum_{k=1}^l\nnz(\mD^{(k)})$.

For example, consider the Metaclust50 matrix in Table~\ref{tab:dataset}.
If we need $r=$24 bytes to store a nonzero (16 bytes for row and column indices and 8 bytes for the value), we need 24TB memory to store the final output. 
However, the actual memory requirement could be up to $92~ \text{Trillion}*24=2208$\revision{TB} (\revision{2.2PB}) as squaring this matrix requires 92 trillion $\flop$.
Cori supercomputer used in our experiment has 1.09PB aggregated memory.
Hence, we need algorithmic innovations to multiply matrices at this extreme scale.

{\bf Batched 3D Sparse SUMMA.}
\label{sec:summa3dbatch}
When the memory requirement of \textproc{Summa3D} exceeds the available memory, we multiply matrices in $b$ batches where each batch computes $n/b$ columns of $\mC$. 
Hence, in a batch, we multiply $\mA {\in} \mathbb{R}^{n \times n}$ and ${\mB {\in} \mathbb{R}^{n \times \frac{n}{b}}}$ to generate ${\mC {\in} \mathbb{R}^{n \times \frac{n}{b}}}$. 
We chose to form batches of $\mC$ column-by-column so that an application can take an informed decision based on the entire columns of $\mC$.
Our motivation comes from graph analytics where we need the entire result for a batch of vertices before an application decides how to process the result and move to the next batch.
For example, Markov clustering~\cite{Azad2018} keeps top-k entries in each column from the resultant matrix.
Hence, we did not consider partitioning C into square tiles despite the fact that it may reduce communication even further.


        

\begin{algorithm}[t]
\caption{Symbolic step to determine $b$.}
{\bf Input}: $\mat{A}$ and $\mat{B}$ are distributed on a 3D process grid $P_{3D}$. 
$M$ denotes the total available memory in bytes and $r$ denotes the number of bytes needed to store each nonzero.
\begin{algorithmic}[1]
\Procedure{Symbolic3D}{$\mat{A}, \mat{B}, P_{3D}, M$}
\For{all processes in $P_{3D}(i,j, k)$ {\bf in parallel}}
    \State $\nnz[i,j,k] \gets 0$ \Comment{per-process nnz}
     \State stages $\gets $ number of process columns in $P_{3D}$
     \For{ $s \gets 1$ to stages} \Comment{SUMMA stages}
     \State $\mat{\Tilde{A}}_{\text{recv}}\gets \Call{Bcast}{\mat{\Tilde{A}}, P_{3D}(i, s, k), P_{3D}(i, :, k)}$
     \State $\mat{\Tilde{B}}_{\text{recv}}$  $\gets$ $\Call{Bcast}{\mat{\Tilde{B}}, P_{3D}(s, j, k), P_{3D}(:, j, k)}$
     \vspace{2pt}
    \State $\nnz[i,j,k] $  += \Call{LocalSymbolic}{$\mat{\Tilde{A}}_{\text{recv}}$, $\mat{\Tilde{B}}_{\text{recv}}$}
    
    \EndFor
\EndFor
\State maxnnzC $\gets$ \Call{AllReduceMax}{$\nnz[i,j,k],  P_{3D}$}
\State maxnnzA $\gets$ \Call{AllReduceMax}{$\nnz(\mat{\Tilde{A}}),  P_{3D}$}
\State maxnnzB $\gets$ \Call{AllReduceMax}{$\nnz(\mat{\Tilde{B}}),  P_{3D}$}
\State $b \gets \frac{r\times \text{maxnnzC}}{M/p-r(\text{maxnnzA+maxnnzB})}$ 
\State \Return $b$
\EndProcedure
\end{algorithmic}
\label{algo:symbolic}
\end{algorithm}

{\bf A symbolic step to determine the required number of batches.}
If $M$ denotes the aggregated memory in $p$ processes, then an SpGEMM algorithm needs at least $b$ phases as follows:
\begin{equation}
    b \geq \ceil[\bigg]{ \frac{\text{mem}(\mC)}{M - r(\nnz(\mA) + \nnz(\mB))} }.
\end{equation}
In practice, $b$ cannot be determined analytically because $\text{mem}(\mC)$ is not known in advance.
$b$ also depends on the layout of the the process grid $P_{3D}$ and the load balancing factor in distributing the matrices. 

We developed the \textproc{Symbolic3D} function (Algorithm~\ref{algo:symbolic}) that estimates the number of phases from input matrices and the process grid $P_{3D}$.
Similar to the \textproc{SUMMA3D} function, \textproc{Symbolic3D} operates in several stages within each layer (line 5-8 in Algorithm~\ref{algo:symbolic}). 
At each stage, local submatrices $\mat{\Tilde{A}}$ and $\mat{\Tilde{B}}$ are broadcast along process rows and columns, respectively. 
After $P_{3D}(i,j,k)$ receives submatrices of $\mA$ and $\mB$ from other processes, it performs a symbolic multiplication locally using \textproc{LocalSymbolic} that computes the number of nonzeros in the output. 
At the end of all SUMMA stages, we find the maximum nnz for $\mA$, $\mB$, and $\mD$ stored at any process (lines 9-11 in Algorithm~\ref{algo:symbolic}).
Finally, line 12 computes $b$ from per-process available memory $M/p$ and other memory requirements.  
Note that the \textproc{Symbolic3D} function considers the maximum unmerged nonzeros stored by a process so that no process exhausts its available memory. 
 This choice makes our algorithm robust to different sparsity patterns.
Hence, in comparison to perfectly-balanced computation, \textproc{Symbolic3D} will estimate more batches for load-imbalanced cases.

Unlike \textproc{SUMMA3D}, \textproc{Symbolic3D} has lightweight computations because \textproc{LocalSymbolic} can be computed much faster than  \textproc{LocalMultiply}. 
By contrast, \textproc{Symbolic3D} still needs expensive broadcasts as was needed by \textproc{SUMMA3D}.
Consequently, the communication-avoiding scheme used in  \textproc{Symbolic3D}  has more significant impact on its performance.



\subsection{The batched SUMMA3D algorithm}

{\bf Data distribution.} 
We perform batching on top of 3D distribution discussed in Fig.~\ref{fig:data_distribution}.
Since a batch computes $n/b$ columns of $\mC$, batching does not change the distribution of $\mA$ shown in Fig.~\ref{fig:data_distribution}.
Globally, a batch should have $n/b$ columns of $\mB$ and $\mC$ across all processes. 
This can be achieved by simply splitting all local submatrices $\Tilde{\mB}$ along the column and making $b$ pieces. 
Let $\Tilde{\mB}[i]$ be the part of local $\mB$ in the $i$th batch.
Since $\Tilde{\mB}$ has $n/\sqrt{p/l}$ columns in 3D distribution, $\Tilde{\mB}[i]$ has $ n/(b\sqrt{p/l})$ columns. 
However, a block column partitioning of $\Tilde{\mB}$ may create load imbalance in the Merge-Fiber operation (to be explained next).
Hence, we use block-cyclic partitioning where each block has $n/(bl\sqrt{p/l})$ columns. 
A collection of $l$ such blocks gives us a batch of desired shape as shown in Fig.~\ref{fig:data_distribution}(i).



\begin{algorithm}[t]
\caption{3D memory-constrained sparse SUMMA with batching.}
\textbf{Input and Output:} Input matrices $\mat{A}$ and $\mat{B}$ and output matrix $\mat{C}$ are distributed in a 3D process grid $P_{3D}$. $\mat{\Tilde{A}}$, $\mat{\Tilde{B}}$, and $\mat{\Tilde{C}}$ denote local submatrices in the current process. 
$b$ is the number of batches.
$\mat{B}[batch]$ denotes a 3D matrix consisting of local batch $\mat{\Tilde{B}}[batch]$ from all processes.
\begin{algorithmic}[1]

\Procedure{BatchedSumma3D}{$\mat{A}, \mat{B}, P_{3D}$}
 \State $b \gets \Call{Symbolic3D}{\mat{A},  \mat{B}, P_{3D}}$
\For{all processes in $P_{3D}(i,j,k)$ {\bf in parallel}} 
  
\vspace{2pt}
  \State $\mat{\Tilde{B}}[1..b]$ $\gets$ ColSplit$(\mat{\Tilde{B}}, b)$
\For{$batch \gets 1$ to $b$}
    \State $\mat{C}[batch] \gets$ \Call{Summa3D}{$\mat{A}$,  $\mat{B}[batch], P_{3D}$}
\EndFor
\State $\mat{\Tilde{C}} \gets$ ColConcat$(\mat{\Tilde{C}}[1..b])$
\EndFor
\State \Return $\mat{C}$
\EndProcedure

\end{algorithmic}
\label{algo:batched3Dsumma}
\end{algorithm}

{\bf The algorithm.}
The \textproc{BatchedSumma3D} function in Algorithm~\ref{algo:batched3Dsumma} multiplies matrices distributed on $P_{3D}$ and generates the result matrix batch by batch. 
At first, each process splits $\Tilde{\mat{B}}$, its local copy of $\mat{B}$, column-wise into $b$ batches $\Tilde{\mat{B}}[1..b]$ (line 4, Alg.~\ref{algo:batched3Dsumma}).
The exact block cycling splitting is described in the previous paragraph and in Fig.~\ref{fig:data_distribution}(i). 
It effectively creates as many pieces of $\mat{B}$ as the number of batches $b$. 
Then for each $batch$, the algorithm calls \textproc{Summa3D} to multiply $\mat{A}$ and $\mat{B}[batch]$ to get the output matrix $\mat{C}[batch]$ (line 6, Alg.~\ref{algo:batched3Dsumma}). 
At the end of all batches, each process has $b$ pieces of $\Tilde{\mC}[1..b]$.
Here, submatrices in $\Tilde{\mC}[1..b]$ have non-overlapping columns. 
Hence, each process concatenates $\Tilde{\mC}[1..b]$ along the column to form its own piece $\Tilde{\mC}$ of the final result  (line 7, Alg.~\ref{algo:batched3Dsumma}).
The overall result matrix $\mat{C}$ is a collection of these local matrices distributed on $P_{3D}$.
Note that we showed the formation of $\mat{C}$ for completeness of the algorithm. 
In practice, the output $\mat{C}[batch]$ from each batch is pruned or saved to disk by the application.
Fig.~\ref{fig:BatchedSUMMA3D} shows an execution of Algorithm~\ref{algo:batched3Dsumma}.
Note that when $\nnz(\mA){\gg} \nnz(\mB)$, column-wise batching can be expensive. 
However, if inputs are square matrices, we can easily use row-by-row batching on $\mat{B}$ using the same algorithm.

{\bf Major steps.}
\textproc{BatchedSumma3D} uses \textproc{Summa3D} in each batch and \textproc{Summa3D} uses \textproc{Summa2D} in each layer. 
Hence, all major steps in \textproc{Summa3D} and \textproc{Summa2D} are also executed in \textproc{BatchedSumma3D}.
Additionally, \textproc{BatchedSumma3D} uses Alg.~\ref{algo:symbolic} to estimate the number of batches needed for an SpGEMM.
Therefore, \textproc{BatchedSumma3D} has seven major steps: 
(1) Symbolic multiplication to estimate $b$ (once; involve communication and computation), (2) A-Broadcast (once per SUMMA2D stage: along a process row on every layer), (3) B-Broadcast (once per SUMMA2D stage: along a process column on every layer), (4) Local-Multiply (once per SUMMA2D stage: local computation), (5) Merge-Layer (once per SUMMA2D stage: local computation), (6) AllToAll-Fiber (once per batch: communicate in a fiber) (7) Merge-Fiber (once per batch: local computation).


\subsection{Communication complexity}
Among seven major steps in \textproc{BatchedSumma3D}, four steps (Symbolic, A-Broadcast, B-Broadcast, and AllToAll-Fiber) involves communication.
When $b>1$, \textproc{BatchedSumma3D} increases the number of times matrix $\mA$ needs to be re-communicated, making the impact of communication-avoidance even more significant for batched SpGEMM. 
Even though we relied on the communication-avoiding technique developed in a prior work~\cite{Azad2016}, batching makes communication steps more fine-grained and irregular.
To analyze the communication complexity we used the $\alpha-\beta$ model,
where $\alpha$ is the latency constant corresponding to the fixed cost of communicating a message regardless of its size, and $\beta$ is the inverse bandwidth corresponding to the cost of transmitting one word of data. Consequently, communicating a message of $n$ words takes $\alpha + \beta n$ time. 

Table~\ref{tab:summary_comm} shows the bandwidth and latency costs of different steps of \textproc{BatchedSumma3D}.
Communication costs of the symbolic step are similar to A-Broadcast and B-Broadcast, except the fact that the communication cost of the symbolic step does not rely on $b$.
Here, the bandwidth bound for AllToAll-Fiber is rather loose because there is already significant compression of intermediate products expected within (a) local SpGEMM calls, and (b) within each SUMMA layer.
As the number of layers $l$ increases, the effect of this compression diminishes. It is tighter to use $\sum_{k=1}^{l}{\nnz(\mD^{(k)})}/p$ in lieu of $\flop/p$, which is a function that grows slowly with $l$.

Table~\ref{tab:summary_comm} shows that all communication steps are performed at least $b$ times.
Consequently, \textproc{BatchedSumma3D} has higher latency overheads relative to \textproc{Summa3D}.
Since, $\mA$ is communicated $b$ times, the bandwidth cost of A-Broadcast could dominate the overall communication cost, especially for a large value of $b$.
Fortunately, we can increase the number of layers to reduce the overhead of re-communicating $\mA$.

\subsection{Faster local computations using hash tables}
\label{sec:fast-comp}
We developed new algorithms for local multiplication and merging within each process with an aim to utilize special structures due to layering and batching. 
One particular optimization is in the sortedness of results after Local-Multiply, Merge-Layer, and Merge-Fiber.
Since the final output $\mC$ is obtained after Merge-Fiber, we keep the output of Merge-Fiber sorted within each column. 
However, the outputs of Local-Multiply and  Merge-Layer do not need to be sorted within each column because they will eventually be sorted after Merge-Fiber.
To facilitate unsorted outputs in Local-Multiply and  Merge-Layer, we used hash-based local SpGEMM and merging algorithms. By contrast, heap-based algorithms used in prior work~\cite{Azad2016} always keep results sorted.
These modifications can make local computations more than $5\times$ faster for many matrices as we will demonstrate in Fig.~\ref{fig:compare}.

{\bf Sort-free local SpGEMM.} Previous 3D SUMMA algorithm~\cite{Azad2016} used a multithreaded heap-based local SpGEMM algorithm in each process. 
However, the heap algorithm was later replaced by a hybrid algorithm that used either a hash table or a heap to form the $i$th column depending on the compression ratio of the column~\cite{Nagasaka2019}. 
After forming the column, that hybrid algorithm sorted the column in an ascending order of row indices. 
With an aim to keep matrices unsorted, we use a hash-based SpGEMM in our Local-Multiply routine (line 7 of Algo.~\ref{algo:2Dsumma}) because the hash SpGEMM does not require sorted matrices as inputs.
Additionally, we refrain from sorting columns once they are formed.
Thus, our local multiplication uses an ``unsorted-hash" algorithm.
In practice, the unsorted-hash algorithm can be $30\%$-$50\%$ faster than the hybrid algorithm.

{\bf Sort-free hash merging algorithms.}
Previous 2D SUMMA~\cite{Buluc2012} and 3D SUMMA~\cite{Azad2016} algorithms used a heap-based merging algorithm in Merge-Layer and Merge-Fiber routines. 
Observing the benefit of hash SpGEMM algorithms~\cite{Nagasaka2019}, we develop a new hash-based merging algorithm for  Merge-Layer and Merge-Fiber steps.
Given a collection of $l$ matrices, the hash-merge algorithm forms the $i$th column of the merged output from the $i$th columns of all input matrices. 
The merging is done using a hash table that can work with unsorted input and outputs an unsorted output column.
This new ``unsorted-hash-merge" algorithm can be an order of magnitude faster than previous heap-merge algorithm.


\begin{table}[!t]
    \centering
    \caption{\revision{Communication complexity for different steps of \textproc{BatchedSumma3D}}.}

    \begin{tabular}{ l | c c c}
    \toprule
         & A- & B- &  AllToAll-  \\
         & Bcast & Bcast  & Fiber \\   
        \toprule
        Per process data & $\nnz(\mA)/p$ & $\nnz(\mB)/(bp)$ &  $\flop/(bp)$\\
        Comm. size & $\sqrt{p/l}$ & $\sqrt{p/l}$  & $l$  \\
        Latency cost & $\alpha\lg{p/l}$ & $\alpha\lg{p/l}$ &  $\alpha l$  \\
         Bandwidth cost & $\beta(\nnz(\mA)/p)$ & $\beta(\nnz(\mB)/(bp))$ &  $\beta(\flop/(bp))$  \\
         How many times & $b(\sqrt{p/l})$ & $b(\sqrt{p/l})$ & $b$  \\
         \midrule
         Tatal latency & $\alpha b(\sqrt{p/l}\lg{p/l})$ & $\alpha b(\sqrt{p/l}\lg{p/l})$ & $\alpha bl$  \\
         Total bandwidth & $\beta b(\nnz(\mA)/\sqrt{pl})$ & $\beta(\nnz(\mB)/\sqrt{pl})$ & $\beta (\flop/p)$  \\
         \bottomrule
    \end{tabular}
    \vspace{-8pt}
    \label{tab:summary_comm}
\end{table}

\begin{table}[!t]
    \centering
    \caption{\revision{Computational complexity for different steps of \textproc{BatchedSumma3D}.}}
    \begin{tabular}{ l |  c c c}
    \toprule
         & Local- & Merge- & Merge   \\
         & Multiply & Layer & Fiber  \\ 
    \toprule
         Complexity &  $\flop/(bp\sqrt{p/l})$ & $\flop/(bp)\lg p/l$  & $\flop/(bp)\lg l$  \\
         How many times & $b(\sqrt{p/l})$ & $b$ & $b$  \\
         \midrule
         Total &  $\flop/p$ & $\flop/p\lg p/l$  & $\flop/p\lg l$  \\

         \bottomrule
    \end{tabular}
    \vspace{-8pt}
    \label{tab:summary_comp}
\end{table}
\begin{table}[!t]
    \centering
    \caption{Overview of the evaluation platform.}
    \begin{tabular}{r l l}
    \toprule
    & \textbf{Cori-KNL} & \textbf{Cori-Haswell} \\
    \toprule
    Processor & Intel Xeon Phi  7250 
& Intel Xeon E5-2698
\\
 
    Cores/node & 68 & 32\\
    Clock & 1.4 GHz & 2.3 GHz \\
    Hyper-threads/core & 4 & 2 \\
    Memory/node & 112GB & 128GB\\

    \toprule

    Total nodes & 9,668 & 2,388\\ 
    Total memory & 1.09 PB & 298.5 TB\\
    
    Interconnect & \multicolumn{2}{l}{Cray Aries with Dragonfly topology} \\
    \toprule
    Compiler &  \multicolumn{2}{l}{Intel icpc Compiler 19.0.3 with -O3 option}   \\
    \bottomrule
    \end{tabular}
    
    \label{tab:system_info}
\end{table}

\begin{table}[!t]
    \centering
    \caption{Statistics about test matrices used in our experiments. $\mC{=}\mA\mA^T$ for Rice-kmers and Metaclust20m. For all other cases, $\mC{=}\mA\mA$. M, B and T denote million, billion and trillion, respectively.
    }
    \begin{tabular}{l r r r r r}
    \toprule
    Matrix ($\mA$) & rows & columns & $\nnz(\mA)$ & $\nnz(\mC)$ & $\flop$ \\
    \toprule
    Eukarya & 3M & 3M & 360M & 2B & 134B \\
    Rice-kmers  & 5M & 2B & 4.5B & 6B & 12.4B \\
    Metaclust20m & 20M & 244M & 2B & 312B & 347B \\
    Isolates-small & 35M & 35M & 17B & 248B & 42T \\
    Friendster & 66M & 66M & \revision{3.6B} & 1T & 1.4T \\
    
    Isolates & 70M & 70M & 68B & 984B & 301T \\
    Metaclust50 & 282M & 282M & 37B & 1T & 92T \\

    \bottomrule
    \end{tabular}
    
    \label{tab:dataset}
\end{table}

\section{Results}
\subsection{Evaluation platforms} 
We evaluate the performance of our algorithm on NERSC Cori system. We used two  types of compute nodes on Cori as described in Table.~\ref{tab:system_info}. 
We used MPI+OpenMP hybrid parallelization.
All of our experiments used 16 and 6 threads per MPI process on Cori-KNL and Cori-Haswell, respectively. 
Only one thread in every process makes MPI calls.


\subsection{Test problems} 
Table~\ref{tab:dataset} describes several large-scale matrices used in our experiments. 
Eukarya, Isolates-small, and Isolates are protein-similarity networks generated from isolate genomes in the IMG
database~\cite{chen2016img}. These matrices are publicly available~\cite{Azad2018}.
Metaclust50 stores similarities of proteins in Metaclust50 (https://metaclust.mmseqs.com/) dataset which contains predicted genes from metagenomes and metatranscriptomes of assembled contigs from IMG/M and NCBI. 
Friendster represents an online gaming network available in the SuiteSparse Matrix Collection~\cite{Davis2011}.
The rows of the Rice-kmers matrix represent the PacBio sequences for the Oryza sativa rice, and its columns represent a subset of the $k$-mers (short nucleotide subsequences of length $k$) that are used by the sequence overlapping software BELLA~\cite{guidi2018bella}. 
The Metaclust20m matrix is originally used in distributed protein similarity search~\cite{sc20pastis} and contains a subset of sequences for Metaclust50.
The rows and columns of this matrix respectively represent protein sequences and k-mers.
The $\mA \mA\transpose$ operation involving this matrix produces candidates for batch pairwise sequence alignment.

Even though there are many applications of SpGEMM, we primarily focus on problems where $\nnz(\mC) > (\nnz(\mA) + \nnz(\mB))$ because for these problems, batching could be needed. 
For example, square of a sparse matrix often requires significantly more memory than input matrices~\cite{Ballard2013}.  
We selected matrices in Table~\ref{tab:dataset} considering the following three applications. 
(a) Markov clustering~\cite{Dongen2000, Azad2018} iteratively squares a protein-similarity matrix to discover protein clusters. We capture this application by extensively studying the computation of $\mA\mA$. 
(b) Clustering coefficients of nodes in a social network are computed by counting triangles. High-performance triangle counting algorithms rely on the multiplication of the lower triangular and upper triangular parts of the adjacency matrix~\cite{Azad2015}. 
$\mA\mA$ captures the computation needed in triangle counting algorithms.
(c) By computing $\mA \mA\transpose$, the sequence overlapper BELLA discovers all (if any) the shared k-mers between all pairs of sequences. By using the sparse matrix as an index data structure in lieu of a hash table, SpGEMM computes this seemingly all-pairs computation without incurring quadratic cost in the number of reads.  

\begin{figure}[!t]
    \centering
    \includegraphics[width=.95\linewidth]{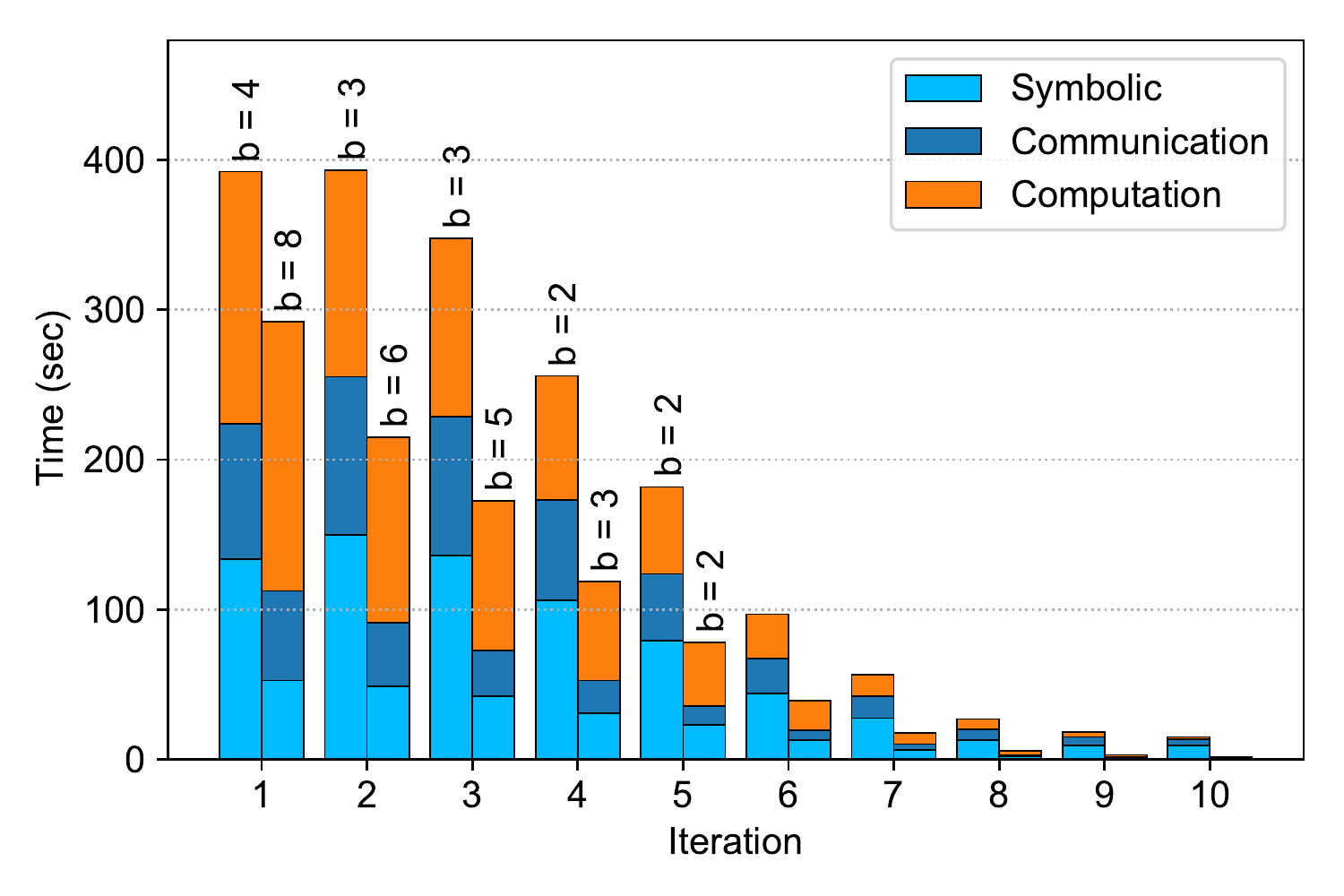} 
    \vspace{-5pt}
    \caption{ Run times of the first 10 iterations of HipMCL when clustering the Isolates-small graph on 65,536 cores of Cori. The left bar of each group represents 1-layer setting and right bar represents 16-layer setting. Number of batches $b$ is calculated using the symbolic step. $b=1$ is not shown. HipMCL needed 66 iterations to cluster proteins of Isolates-small graph. Overall, it was $1.88\times$ faster with 16 layers than with 1 layer.}
    \vspace{-10pt}
    \label{fig:hipmcl-isolates-small}
\end{figure}
\subsection{Impact on an end-to-end protein clustering application}
We plugged the newly developed SpGEMM algorithm in HipMCL~\cite{Azad2018}, a distributed-memory implementation of the Markov clustering algorithm.
HipMCL clusters a protein-similarity network by iteratively squaring the input matrix and then applying column-wise pruning to keep the output sparse.
For large networks in Table~\ref{tab:dataset}, Cori does not have enough memory to store $\mA^2$.
When we employ \textproc{BatchedSumma3D} inside HipMCL, we form $\mA^2$ batch-by-batch, apply various pruning strategies on the current batch and then proceed to the next batch.
Hence, the presented algorithm satisfies the need of HipMCL perfectly well.

Fig.~\ref{fig:hipmcl-isolates-small} show that the runtime of the first 10 iterations of HipMCL when \textproc{BatchedSumma3D} is used with 1 and 16 layers. 
In the first few iterations, HipMCL performs expensive multiplications that needs multiple batches on 1024 nodes. 
Even though the 16-layer setting needs more batches, the benefit of communication avoidance makes most iterations $2\times$ or more faster than 2D SpGEMM. 
Overall, \textproc{BatchedSumma3D} makes HipMCL more than $1.88\times$ faster than a batched 2D algorithm. 
More importantly, {\em HipMCL cannot even cluster Isolates-small on 1024 nodes of Cori if batching is not used}. 
Hence, the presented algorithm makes large-scale protein clustering possible, and at the same time, it makes HipMCL significantly faster than previous algorithms.

\begin{figure*}
    \centering 
    \includegraphics[width=0.315\linewidth]{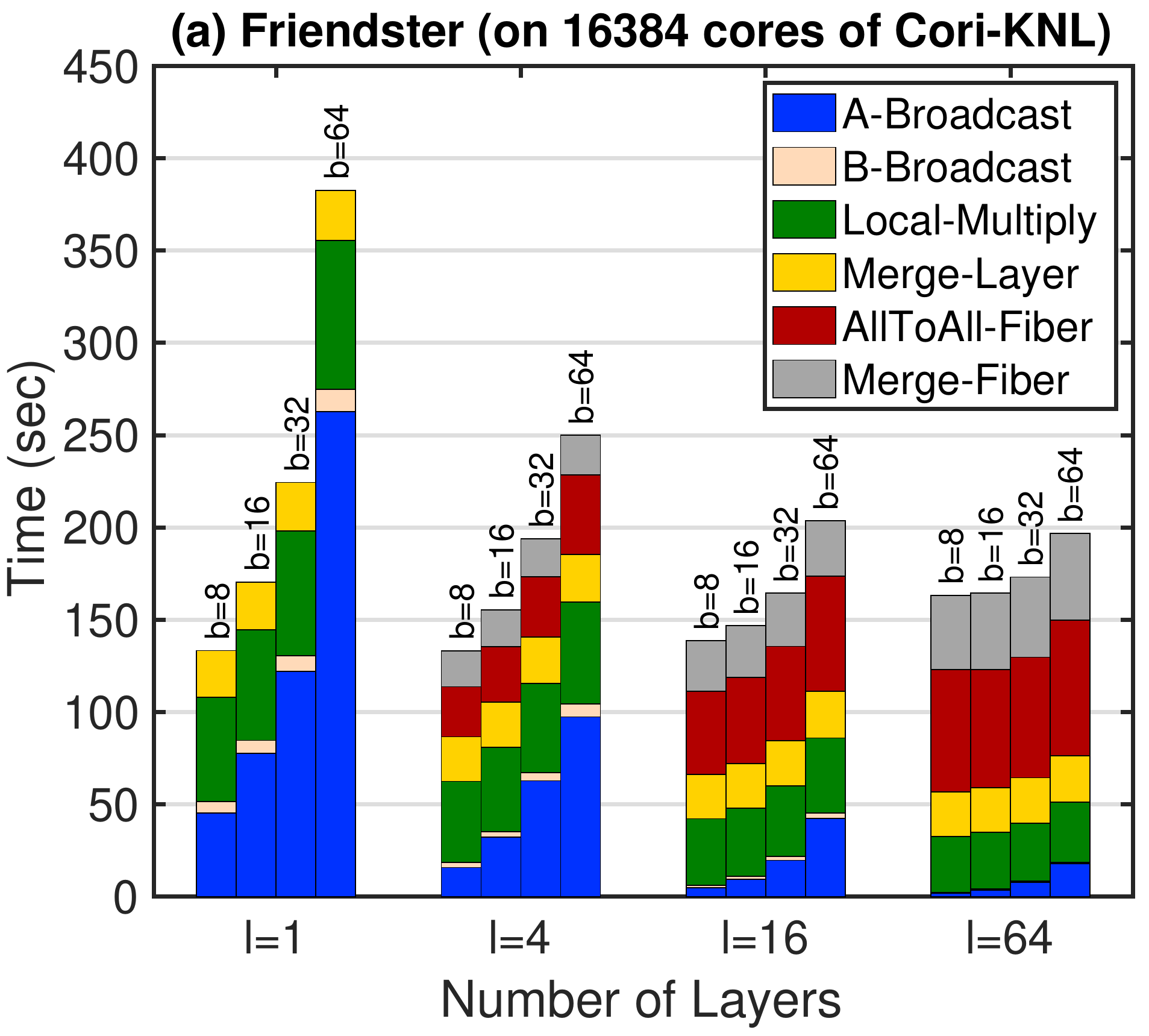} 
    \includegraphics[width=0.345\linewidth]{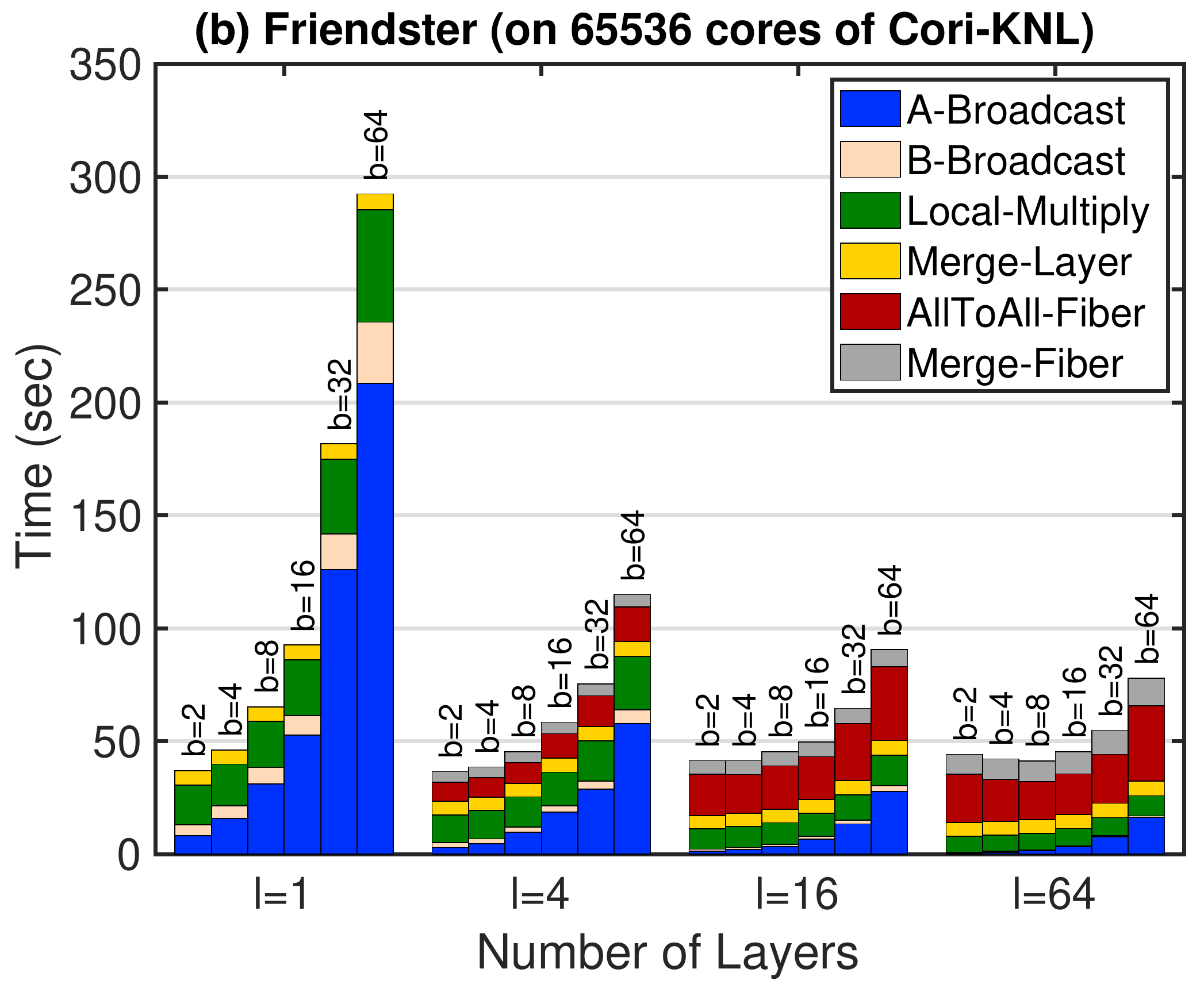}
    \includegraphics[width=0.325\linewidth]{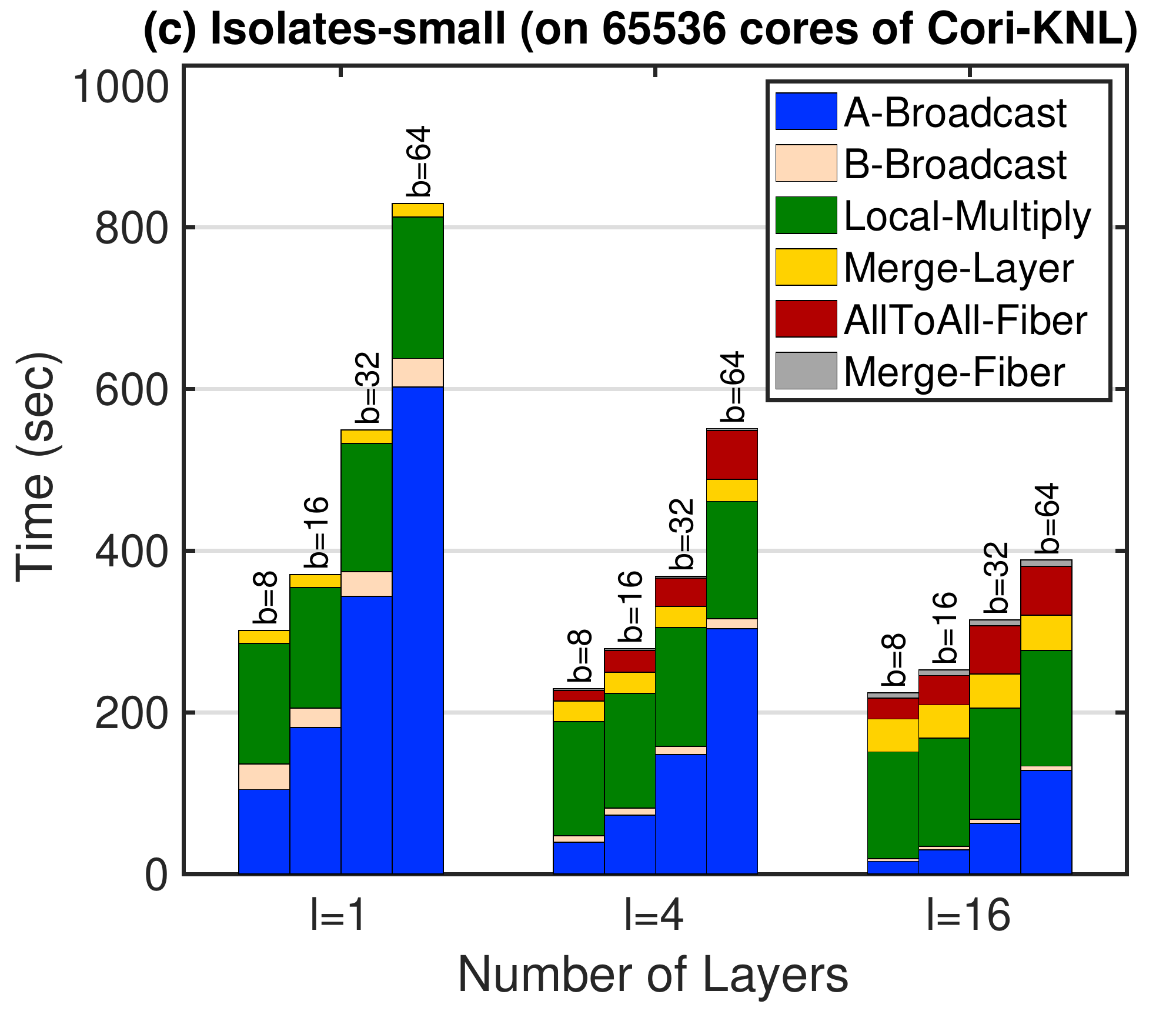}
    \caption{Evaluating the impact of the number of layers (denoted by $l$) and batches (denoted by $b$) on different steps of \textproc{BatchedSumma3D}. In each experiment, a fixed number of nodes and 16 threads per process are used. Runtimes shown in a bar denote the total time needed for all batches shown at the top of the bar. (a) Squaring Friendster on 16,384 cores (256 nodes and 1024 processes); (b) Squaring Friendster on 65,536 cores (1024 nodes and 4096 processes); (c) Squaring Isolates-small on 65,536 cores (1024 nodes and 4096 processes). 
    }
    \vspace{-12pt}
    \label{fig:benchmark}
\end{figure*}

\subsection{Evaluating the impact of number of layers and batches}
Given the use of matrix squaring in HipMCL, we use the computation of $\mA^2$ to show various feature of our algorithm.
At first, we study the impact of $l$ and $b$ on different steps of \textproc{BatchedSumma3D}.
Even though a suitable value of $b$ depends on the available memory and is determined by the symbolic step, we vary $b$ in this experiment to capture different amount of memory available in different distributed systems. 
Fig.~\ref{fig:benchmark} shows the results with two matrices and two different number of cores. 
We summarize key observations from Fig.~\ref{fig:benchmark}.

\begin{figure}[!t]
    \centering
    \includegraphics[width=.85\linewidth]{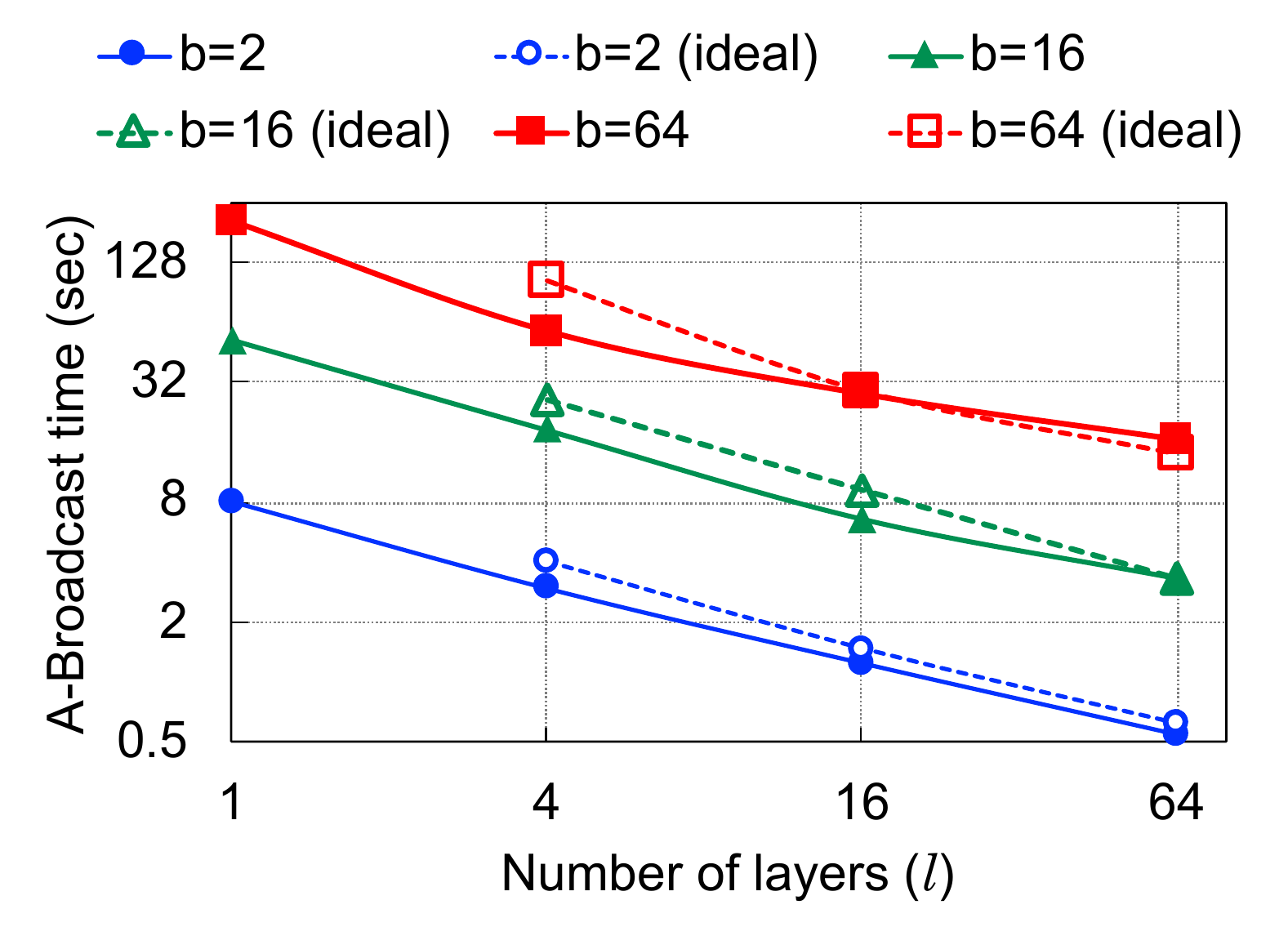} 
    \vspace{-8pt}
    \caption{With a fixed $b$, A-Broadcast time decreases when $l$ increases. Here, we multiply Friendster on 65,536 cores (1024 nodes) with different batches (come from Fig.~\ref{fig:benchmark}(b)). Solid lines denote  observed A-Broadcast time and  dashed lines denote expected A-Broadcast time that decreases by a factor of 2 as we increase $l$ by a factor of 4. 
    }
    \vspace{-10pt}
    \label{fig:Abcast_bench}
\end{figure}

{\bf A-Broadcast time:}
{\em With a fixed $l$, A-Broadcast time increases almost linearly with $b$}. This is expected as $\mA$ is broadcast $b$ times in total (once in every batch).
{\em With a fixed $b$, when $l$ increases, A-Broadcast time decreases at a rate proportional to $\sqrt{l}$}. 
Fig.~\ref{fig:Abcast_bench} shows this observation for different batches when multiplying Friendster on 65,536 cores.
In Fig.~\ref{fig:Abcast_bench}, solid lines (observed A-Broadcast time) closely follow dashed line (expected based on a factor of $\sqrt{l}$ decrease).
To explain this, consider a $\sqrt{p/l}\times\sqrt{p/l} \times l$  process grid with $p$ processes.
If we increase the number of layers by a factor of 4, the grid becomes $\frac{1}{2}\sqrt{p/l}\times \frac{1}{2}\sqrt{p/l} \times 4l$.
Hence, the number of processes in each process row within a layer is reduced by a factor of 2. 
Since A-Broadcast is performed within individual process rows of each layer, the communicator size for A-Broadcast would be decreased by a factor 2.  
Thus, the A-Broadcast time decreases at a rate proportional to $\sqrt{l}$.

{\bf B-Broadcast time:} {\em With a fixed $l$, the B-Broadcast time does not change with $b$}. This is expected because the total data volume associated with B-Broadcast remains the same with batching. Hence, the bandwidth cost for B-Broadcast does not depend on $b$. However, the latency term increases linearly as we increase $b$. Since small latency-bound matrices may not need batching, B-Broadcast time does not rely on $b$ as observed in Fig.~\ref{fig:benchmark}. 
{\em With a fixed $b$, B-Broadcast time decreases when we increase $l$}. This observation is consistent with the corresponding observation in  A-Broadcast.

\begin{figure*}[!t]
    \centering
    \includegraphics[width=0.33\linewidth]{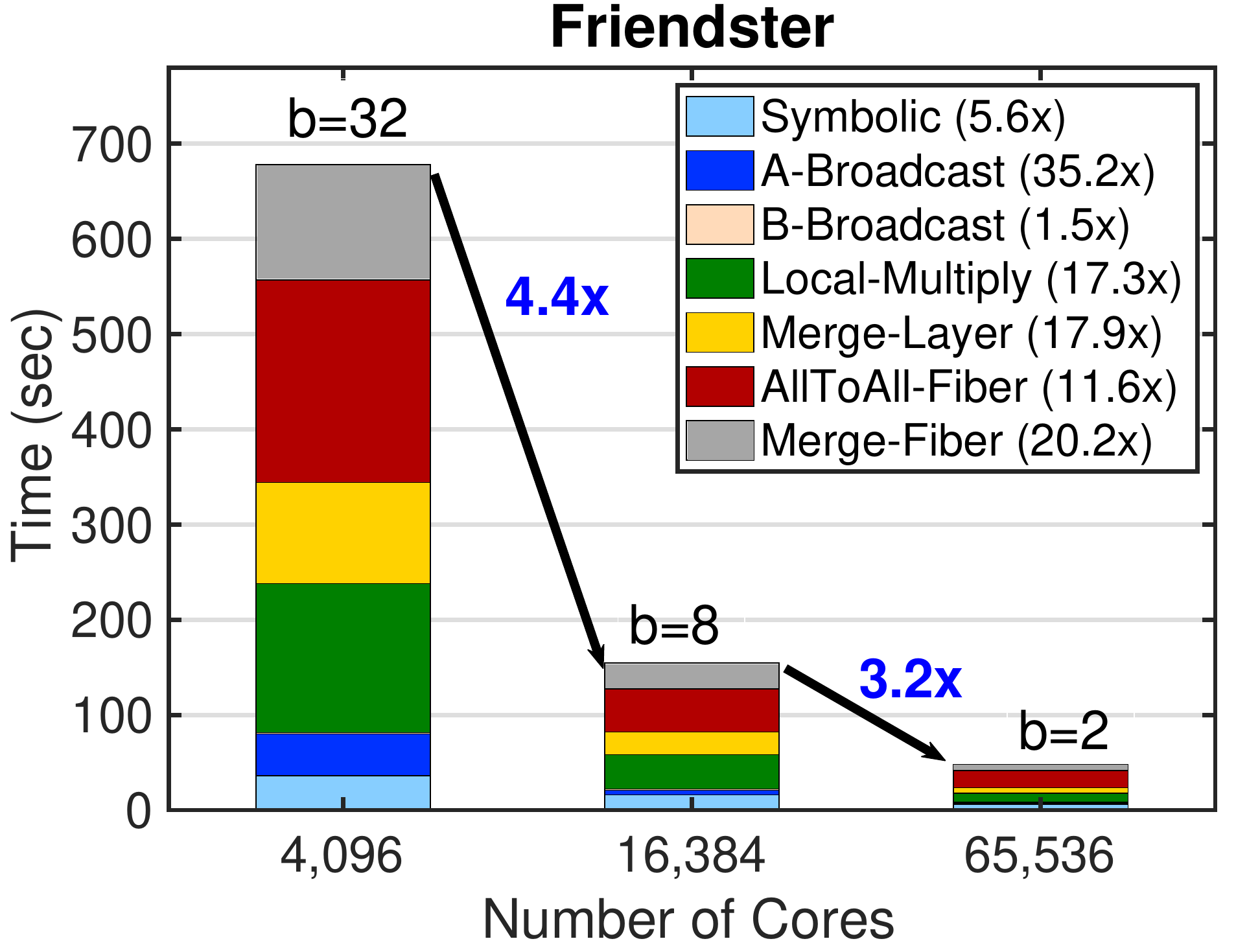} \quad
    \includegraphics[width=0.33\linewidth]{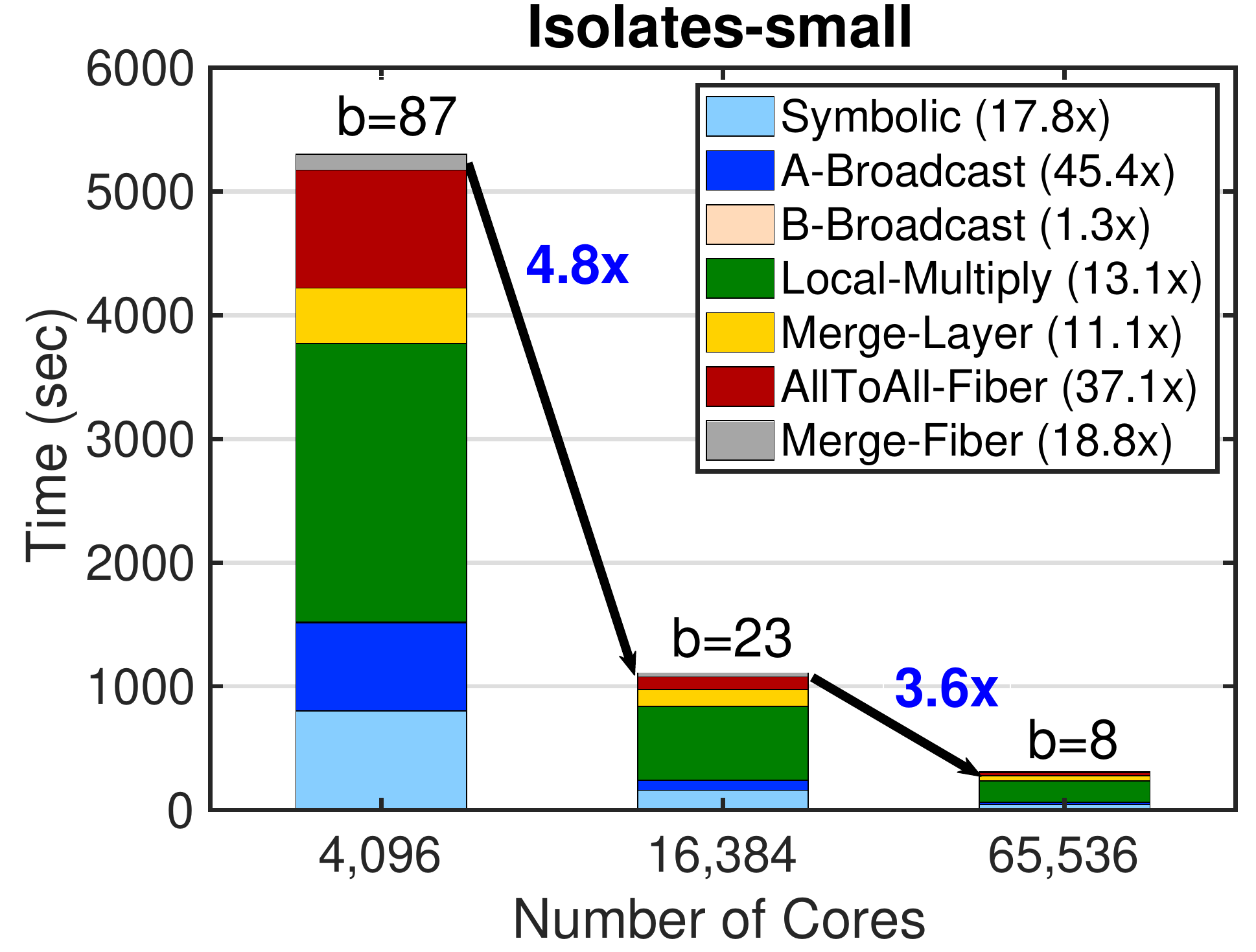}
    \vspace{-2pt}
    \caption{Strong scaling when squaring Friendster and Isolates-small. The scaling experiments were conducted from 64 nodes (4,096 cores) to 1024 nodes (65,536 cores) on Cori-KNL. Number of layers is set to 16 and number of batches is shown on top of each bar.  Total speedups from one bar to the next bar are shown by arrowheads. Numbers in captions denote the overall speedup of each step in batched-SUMMA3D when we go from 4,096 to 65,536  cores (a 16$\times$ increase in core counts). }
    \vspace{-5pt}
    \label{fig:scaling-small}
\end{figure*}
\begin{figure*}[!t]
    \centering
    \includegraphics[width=0.33\linewidth]{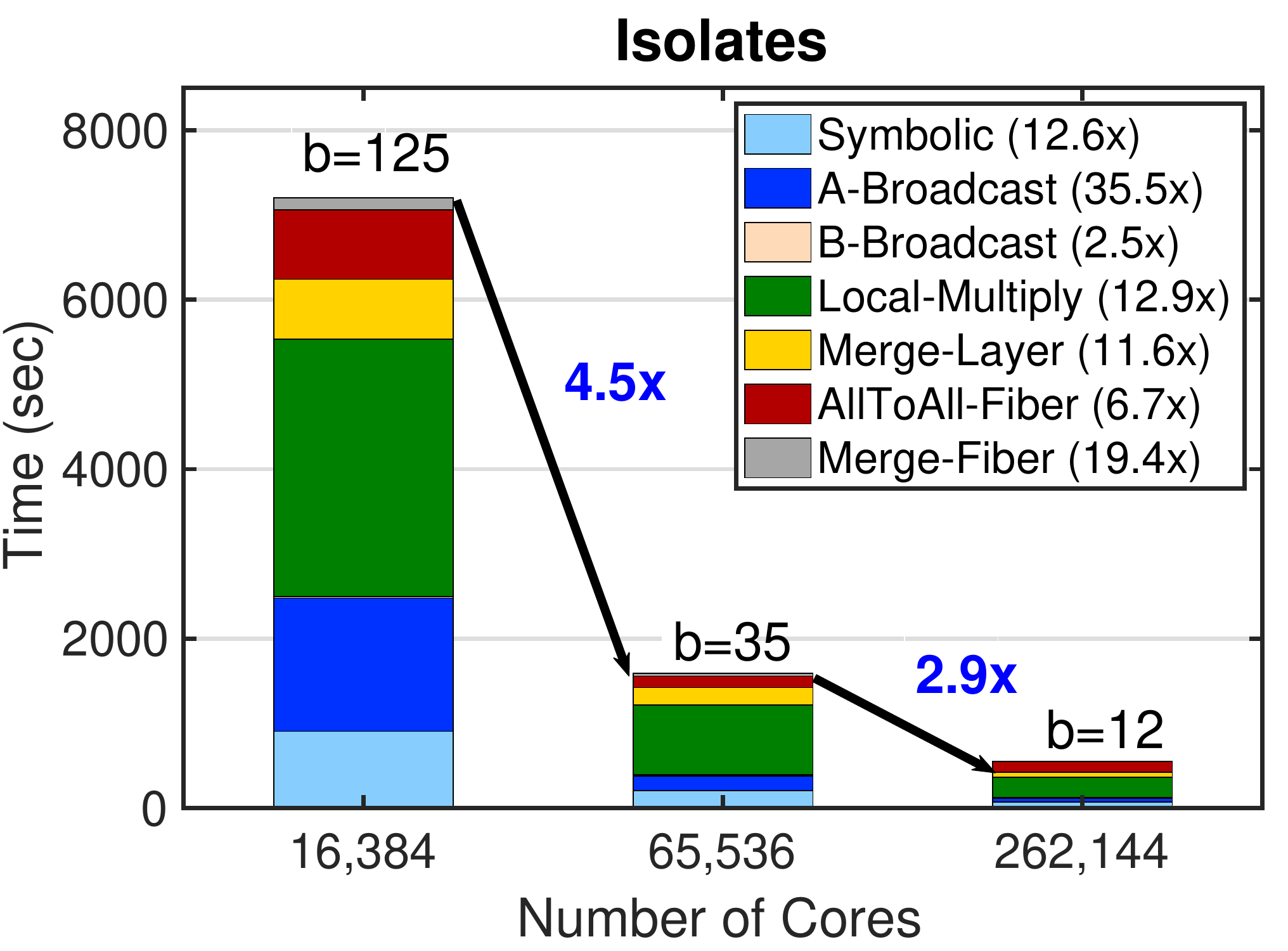} \quad
    \includegraphics[width=0.33\linewidth]{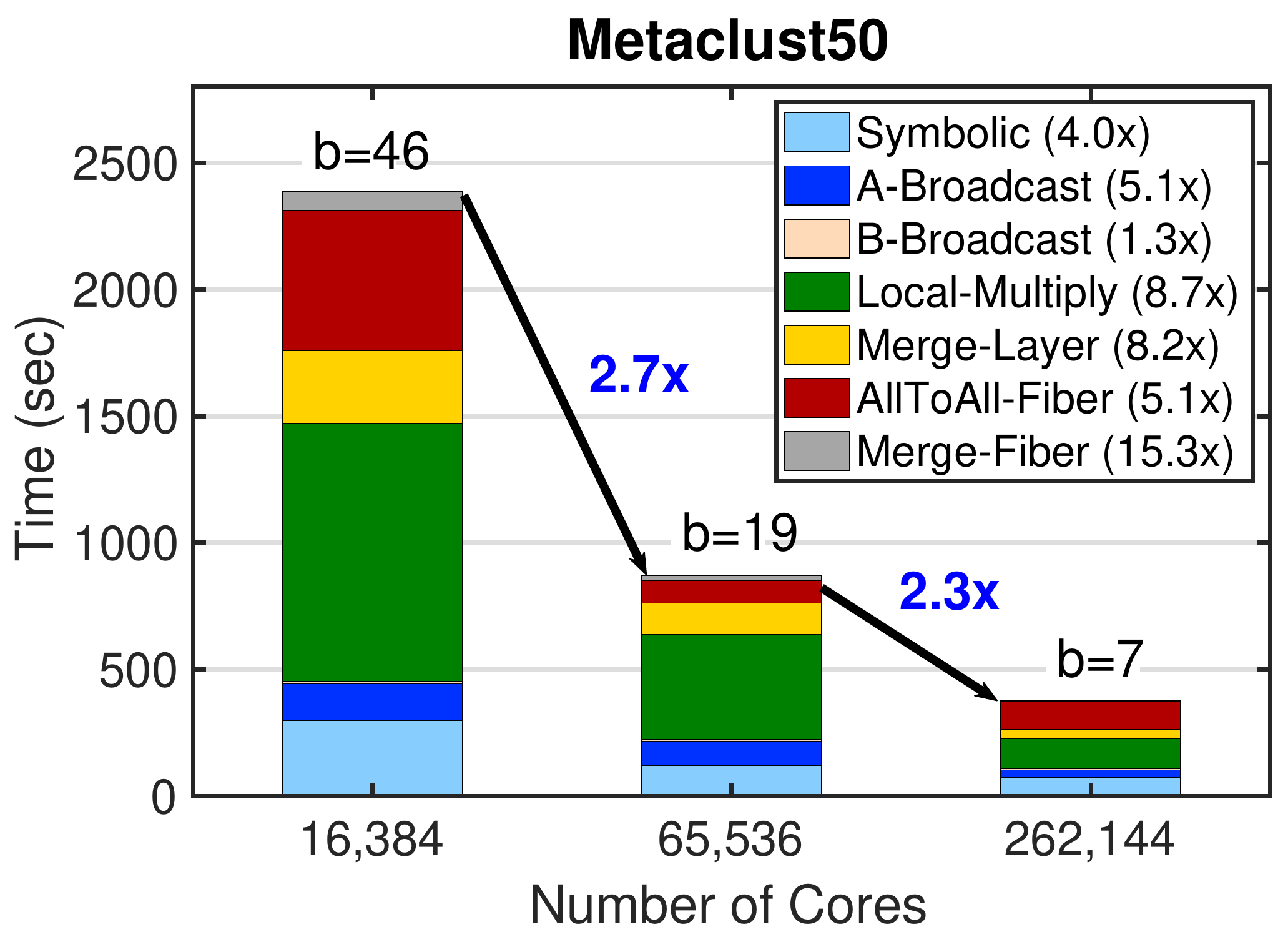}
    \vspace{-5pt}
    \caption{Strong scaling when squaring two biggest matrices in our test suite. The scaling experiments were conducted from 256 nodes (16,384 cores) to 4096 nodes (262,144 cores) on Cori-KNL. Number of layers is set to 16 and number of batches is shown on top of each bar. Total speedups from one bar to the next bar are shown by arrowheads. Numbers in captions denote the overall speedup of each step when we go from 16,384 to 262,144 cores (a 16$\times$ increase in core counts).}
    \vspace{-8pt}
    \label{fig:scaling-big}
\end{figure*}

{\bf Local-Multiply time: }{\em With a fixed $l$, the Local-Multiply time does not change significantly with $b$}. As we increase $b$, per layer result matrix is computed in increasing number of batches. It does not change the complexity or data access patterns of local multiplication. 
However, if $b$ is significantly large, the repeated cost of thread creation, memory allocation, and cache accesses may increase the Local-Multiply time as observed with $b=64$ in Fig.~\ref{fig:benchmark}. {\em With a fixed $b$, the Local-Multiply time decreases with the increase of $l$.} Suppose in a Local-Multiply with $l=1$, we multiply an $\hat{n}\times \hat{k}$ matrix $\hat{\mA}$ with a $\hat{k}\times \hat{n}$ matrix $\hat{\mB}$ and generate an $\hat{n}\times \hat{n}$ matrix $\hat{\mD}$. With $l$ layers, each layer multiplies an $\hat{n} \times \frac{\hat{k}}{l}$ matrix $\hat{\mA}$ with a $\frac{\hat{k}}{l}\times \hat{n}$ matrix $\hat{\mB}$ and generate an $\hat{n}\times \hat{n}$ matrix $\hat{\mD}$. 
Therefore, as we increase $l$, Local-Multiply generates a lower-rank version of the final output.
As a result, with the increase of $l$, the complexity of local multiplication decreases. 
This effect is more significant for sparser matrices, where the result of local multiplication becomes hyper-sparse with many layers.
For example with $b=64$,  when we go from $l=1$ to $l=16$, Local-Multiply decreases by a factor of $3.6\times$ for Friendster (Fig.~\ref{fig:benchmark}(b)) and by a factor of $1.2\times$ for Isolates-small (Fig.~\ref{fig:benchmark}(c)).

{\bf AllToAll-Fiber time:} {\em With a fixed $l$, the AllToAll-Fiber time does not change significantly with $b$}. For a fixed $l$, AllToAll-Fiber is performed among $l$ processes on each fiber in $b$ batches. If  AllToAll-Fiber in each batch is bandwidth-bound, its runtime does not change with $b$ as is observed in   Fig.~\ref{fig:benchmark}.
{\em With a fixed $b$, the AllToAll-Fiber time increases as we increase $l$.}
As explain before, increasing $l$ creates lower rank output in each layer, needing more data to be communicated across layer for merging.
Furthermore, increasing $l$ also increases the communicator size of AllToAll-Fiber.
Hence, the AllToAll-Fiber time increases with the increase of $l$.

{\bf Merge-Fiber time:} {\em With a fixed $l$, the Merge-Fiber time does not change significantly with $b$.} As with AllToAll-Fiber, the Merge-Fiber time is not influenced by $b$.
{\em With a fixed $b$, the Merge-Fiber time increases as we increase $l$.} As explain in the previous paragraph, increasing $l$ requires more data to be merged in the Merge-Fiber step.

\begin{figure}[!t]
    \centering
    \includegraphics[width=.9\linewidth]{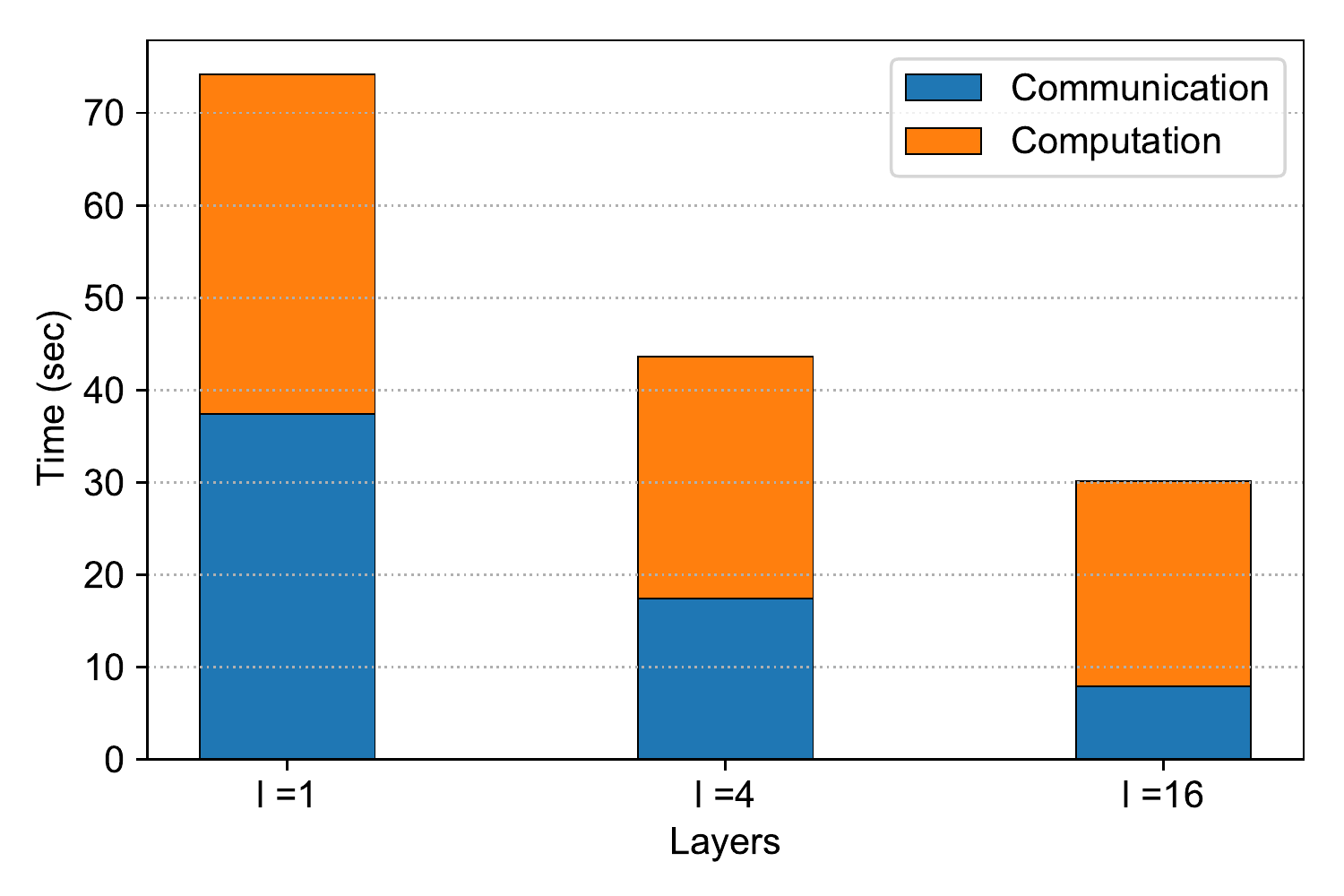} 
    \caption{ Comparing computation and communication time in the symbolic step for the Isolates-small graph on 65,536 Cori KNL cores.
    }
    \vspace{-5pt}
    \label{fig:symbolic-isolates-small}
\end{figure}

{\bf Symbolic time:}
The symbolic step used to determine $b$ also uses a communication-avoiding algorithm as discussed in Algo.~\ref{algo:symbolic}. 
Fig.~\ref{fig:symbolic-isolates-small} shows that the symbolic step becomes significantly faster as we increase $l$.
The communication time in the symbolic step becomes more than $4\times$ faster when we use 16 layers, which results in more than $2\times$ speedup of the total symbolic runtime. 
Here, the communication-avoiding algorithm has a higher impact than the actual multiplication because Algo.~\ref{algo:symbolic} needs lighter local computation.

\begin{table}[!t]
    \centering
    \caption{The impact of $l$ and $b$ on different steps of \textproc{BatchedSumma3D}. $\leftrightarrow$: no change, $\uparrow$: increase, $\downarrow$: decrease. The Local-Multiply time may slightly increase with $b$.}
    \begin{tabular}{c c | c c c c c c}
    \toprule
        & & A- & B- & Local- & Merge- & Merge & AllToAll-   \\
        $l$ & $b$ & Bcast & Bcast & Multiply & Layer & Fiber & Fiber \\   
        \toprule
        $\leftrightarrow$ & $\uparrow$ & $\uparrow$ & $\leftrightarrow$ & \rotatebox[origin=c]{120}{$\downarrow$} & $\leftrightarrow$ & $\leftrightarrow$ & $\leftrightarrow$\\
        $\uparrow$ & $\leftrightarrow$ & $\downarrow$ & $\downarrow$ & $\downarrow$ & $\leftrightarrow$ & $\uparrow$ & $\uparrow$  \\
         \bottomrule
    \end{tabular}
    \vspace{-8pt}
    \label{tab:summary_impact}
\end{table}

{\bf Selecting the number of layers and batches.}
Table~\ref{tab:summary_impact} summarizes the overall impacts of $l$ and $b$ on different steps of \textproc{BatchedSumma3D}.
In the rest of our experiments, we set $b$ to the smallest possible value so that the result in a batch fits in the avilable memory. 
Generally, the A-Broadcast and B-Broadcast time continue to decrease as we increase the number of layers as seen in Fig.~\ref{fig:benchmark}.
However, All2All-Fiber and Merge-Fiber time increase as we increase $l$.
Therefore, selecting the optimum number of layers is challenging as it depends on the tradeoff between broadcasts and fiber reduction/merge costs. 
In the rest of our experiments, we set $l=16$ as it usually gives the best result as can be seen in Fig.~\ref{fig:benchmark}.

\subsection{Strong scaling results}


\vspace{-2pt}
Fig.~\ref{fig:scaling-small} and Fig.~\ref{fig:scaling-big} show the strong scaling of batched-SUMMA3D for four different matrices.
We summarize the strong scaling results with the following key findings.

{\bf \textproc{BatchedSumma3D} scales remarkably well at extreme scale.}
As we increase core counts by $16\times$, our algorithm scales remarkably well for all four matrices: Friendster ($14\times$), Isolates-small ($17.3\times$),  Isolates ($13\times$), and Metaclust50 ($6.3\times$).
For bigger matrices \textproc{BatchedSumma3D} scales to 4096 nodes (262,144 cores) and possibly beyond. 
This scalability is due to two factors: (a) all computations (Local-Multiply, Merge-Layer, and Merge-Fiber) scale almost linearly as shown in the captions of the scalability figures, and (b) dominant communication costs (A-Broadcast and AllToAll-Fiber) also scale well, thanks to the communication-avoiding algorithm. In fact, A-Broadcast can scale super-linearly (e.g., $45.4\times$ reduction for the Isolate-small matrix in Fig.~\ref{fig:scaling-small}) because of the reduced number of batches needed at higher concurrency. The Symbolic step also scales well. Only B-Broadcast does not scale as well as other steps possibly because of high latency overhead in broadcasting small pieces of $\mB$. However, B-Broadcast constitutes about 1\% of the total runtime for all matrices in Fig.~\ref{fig:scaling-small} and Fig.~\ref{fig:scaling-big}.
Hence, the B-Broadcast time does not significantly impact the scalability of \textproc{BatchedSumma3D}.

{\bf Super linear speedup is attainable with more aggregated memory at high node counts.}
As we increase the number of nodes by $4\times$, the aggregated memory also increases by a factor of 4.
As a result, $b$ decreases by at least a factor of $2$ in all cases in Fig.~\ref{fig:scaling-small} and Fig.~\ref{fig:scaling-big}.
Fewer batches at higher concurrency, can super-linearly reduce the A-Broadcast time, resulting in possible super-linear speedups. 
For example, for Isolates in Fig.~\ref{fig:scaling-big}, the total runtime decreases by $4.5\times$ when we go from 256 nodes (16,384 cores) to 1024 nodes (65,536 cores) because  $b$ decreases from 125 to 35.
The super-linear speedup is especially observed at low concurrencies because of the dramatic reduction of the number of batches. 
Even though $b$ decreases as we increase the number of nodes, their relationship is not straightforward as they are related via the intermediate per-layer expanded matrix $\mD^{(k)}$ that in turn depends on per-layer compression factor as well as the overall compression factor. 
For example, in Fig.~\ref{fig:scaling-big}, when we go from 65K cores to 262K cores, the number of batches is decreased by less than 3x even though the memory increases by 4x.

\begin{figure}[!t]
    \centering
    \includegraphics[width=.85\linewidth]{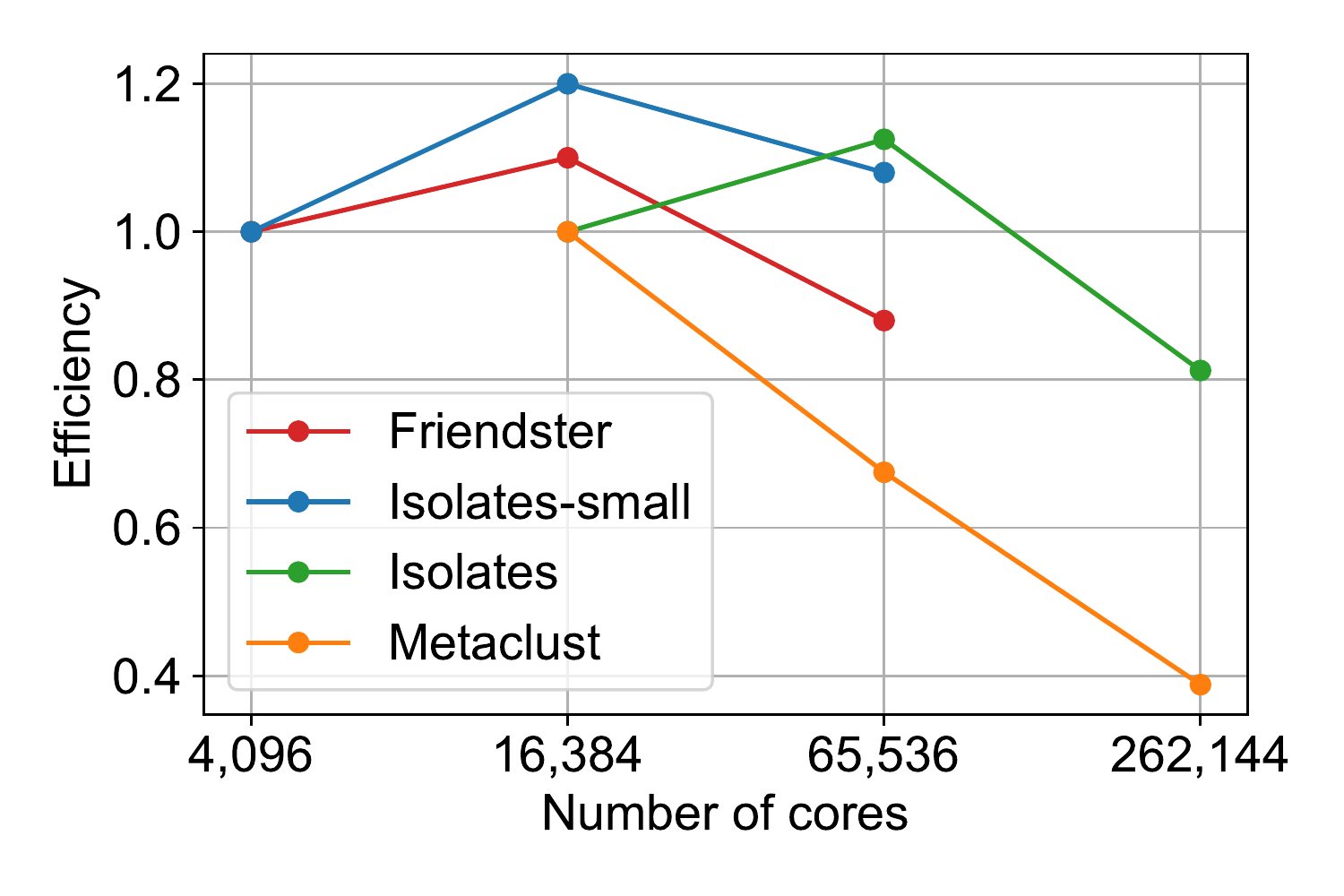} 
    \vspace{-8pt}
    \caption{
    \revision{Parallel efficiency of \textproc{BatchedSumma3D}.}
    }
    \vspace{-10pt}
    \label{fig:efficiency}
\end{figure}

{\bf Parallel efficiency.} We compute the parallel efficiency by using  $\frac{P1}{P2}\frac{T(P1)}{T(P2)}$ where $T(P)$ denotes the runtime with $P$ processes, and $P2{>}P1$. Fig.~\ref{fig:efficiency} shows the parallel efficiency of four large matrices. We observe that the efficiency remains close to 1 for three out of four large matrices. Here, superlinear speedups resulted in an efficiency greater than one. For Metaclust, the efficiency drops to 0.4 on 262K cores. Since  Metaclust is sparser than the other big matrix Isolates, the communication cost for Metaclust  starts to dominate quickly. For example, on 4096 nodes, the communication takes  48\% and 36\% of the total runtime for Metaclust and Isolates, respectively. As communication does not scale as well as computation, we observe a drop in parallel efficiency for sparser matrices like Metaclust.

\begin{table}[!t]
    \centering
    \caption{Overview of local computation improvements when multiplying Isolates-small on 65,536 Cori KNL cores.}
\begin{tabular}{|c | l l |l l| l l |}
  \hline
  \multirow{2}{*}{Layers} 
      & \multicolumn{2}{|c|}{Local-Multiply} 
      & \multicolumn{2}{|c|}{Merge-Layer} 
      & \multicolumn{2}{|c|}{Merge-Fiber} \\             \cline{2-7}
  & Previous & Now & Previous & Now & Previous & Now \\  \hline
  $1$ & 144s & 148s & 258s & 16.5s & - & - \\      
  $4$ & 149s & 135s & 349s & 26.2s & 16.7s & 2.2s \\      
  $16$ & 172s & 130s & 443s & 39.9s & 74.7s & 7.3s \\      \hline
\end{tabular}
\label{tab:computation-improvement}
\vspace{-5pt}
\end{table}

\subsection{The impact of faster computational kernels}
As mentioned in Sec.~\ref{sec:fast-comp}, we developed new hash-based merging algorithms that replaced a heap-based merging algorithms used in previous work~\cite{Azad2016}.
As hash-based SpGEMM and merging algorithm can work with unsorted matrices, we keep all intermediate results unsorted to reduce the computational complexity.
These two modifications (hash-based merging and unsorted matrices) made local multiplication and merging significantly faster as shown in Table~\ref{tab:computation-improvement}.

We observe that the unsorted-hash SpGEMM algorithm used in the Local-Multiply step of \textproc{BatchedSumma3D} can be up to $30\%$ faster than the previous hybrid SpGEMM algorithm when 16 layers are used.
The performance benefit of unsorted local multiplication is more significant with more layers.
This is because the number of nonzeros in intermediate matrices $\mD^{k}$ increases as we increase $l$, which requires sorting with a larger volume of data.
When $l{=}1$, the previous hybrid algorithm can run faster than the unsorted-hash algorithm because the hybrid algorithm can also use a heap-based algorithm that is often faster when the compression ratio of a column is small~\cite{Nagasaka2019}.

Table~\ref{tab:computation-improvement} shows a dramatic improvement when we use the new unsorted-heap-merging algorithm instead of the previous heap-merging algorithm.
On 16 layers, both Merge-Layer and Merge-Fiber time reduces by an order of magnitude.
Consequently, the total computation of \textproc{BatchedSumma3D} is made at least $8\times$ faster than previous state-of-the-art SUMMA 3D implementation~\cite{Azad2016}. 

\begin{figure}[!t]
    \centering
    \includegraphics[width=.8\linewidth]{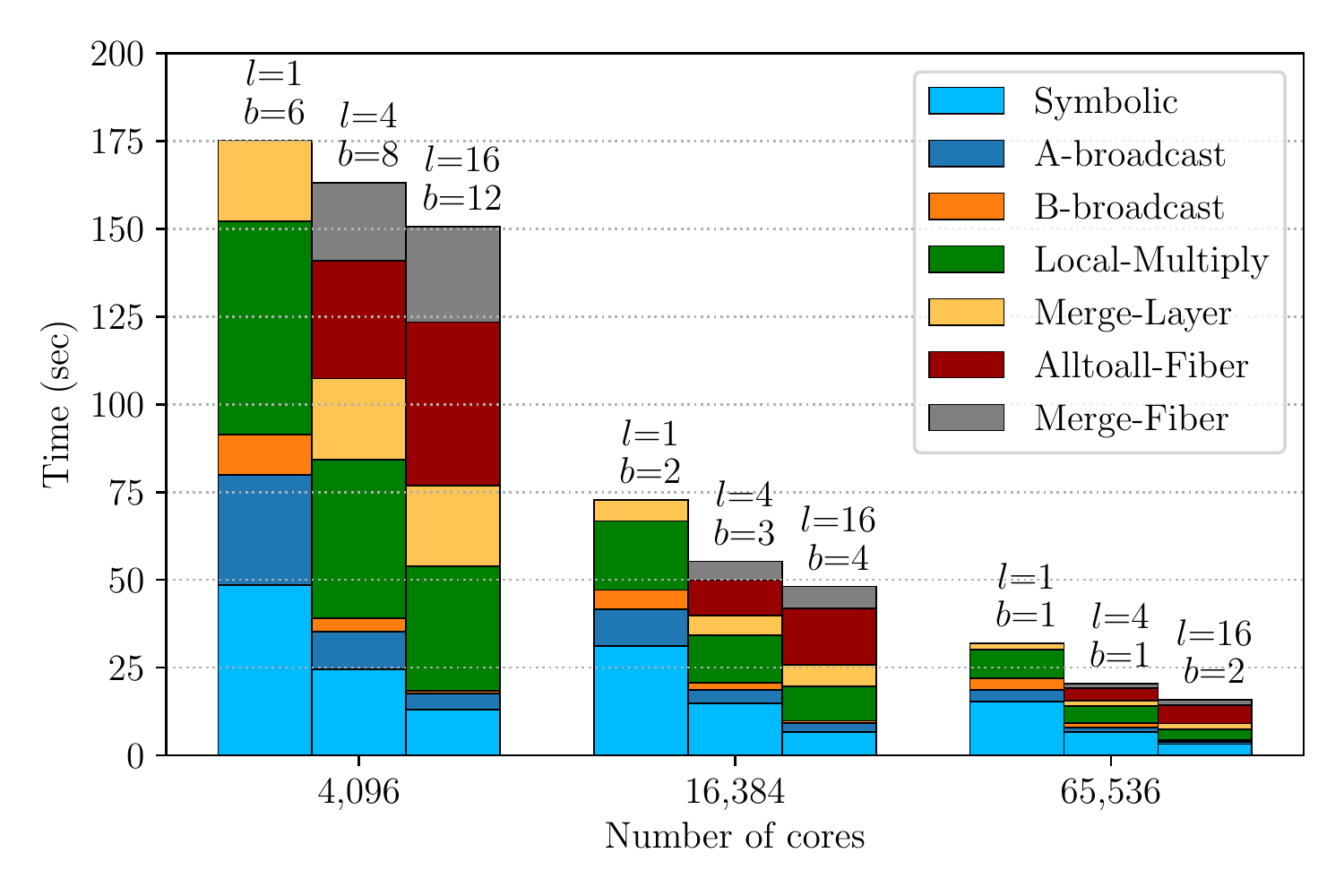} 
    \vspace{-10pt}
    \caption{Computing $\mA\mA\transpose$ with the Metaclust20m matrix on Cori.}
    
    \label{fig:aat-metaclust-20m}
\end{figure}

\begin{figure}[!t]
    \centering
    \includegraphics[width=.7\linewidth]{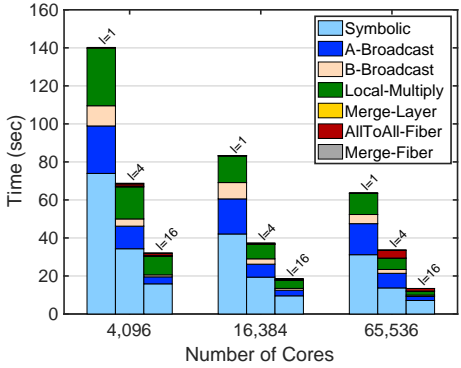} 
    \caption{Scalability of computing $\mA\mA\transpose$ with the Rice-kmers matrix on Cori-KNL.}
    \vspace{-12pt}
    \label{fig:AAT_Scaling}
\end{figure}

\subsection{The scalability when computing $\mA\mA \transpose$}
\vspace{-2pt}
We compute $\mA \mA\transpose$ only for Metaclust20m and Rice-kmers matrices considering their application in sequence overlapper BELLA~\cite{guidi2018bella} in computing the shared k-mers between all pairs of sequences.

Fig.~\ref{fig:aat-metaclust-20m} shows the performance of \textproc{BatchedSumma3D} when computing $\mA \mA\transpose$ for  Metaclust20m.
On 64 nodes (4K cores), \textproc{BatchedSumma3D} with 16 layers is slightly faster than with 1 layer. This is because the 16-layer instance needed 12 phases whereas the 1-layer instance needed just 6 phases. Hence, on 64 nodes, the benefit of communication-avoidance is overshadowed by the need to broadcast $\mA$ $2\times$  more time. 
On 1024 nodes (65K cores), \textproc{BatchedSumma3D} with 16 layers is about $2\times$ faster than with 1 layer even though the 1-layer case does not need any batching. 
The latter case highlights the significant performance benefit of \textproc{BatchedSumma3D} at high concurrencies with or without batching. 

Table~\ref{tab:dataset} shows that $\nnz(\mA \mA\transpose)\approx \nnz(\mA)$ for the Rice-kmers matrix. 
Hence, batching is often not required to compute $\mA \mA\transpose$ with Rice-kmers.
In this case, \textproc{BatchedSumma3D} computes $\mA \mA\transpose$ with $b{=}1$, while the communication-avoiding algorithm reduces the communication costs.
Fig.~\ref{fig:AAT_Scaling} shows that $\mA \mA\transpose$ computation on the Rice-kmers matrix is dominated by communication when only one layer is used (recall that the Symbolic step also performs A-Broadcast and B-Broadcast).
This is expected since Rice-kmers has just 2 nonzeros per column on average. 
As SpGEMM is dominated by communication, using more layers reduces the runtime significantly, as expected. 
For example, on 65,536 cores (1024 nodes) of Cori-KNL, $\mA \mA\transpose$ can be computed $6\times$ faster if we use 16 layers instead of 1 layer.  
This experiment demonstrates that \textproc{BatchedSumma3D} helps any SpGEMM run faster at extreme scale with our without batching.

\begin{figure}[!t]
    \centering
    \includegraphics[width=.88\linewidth]{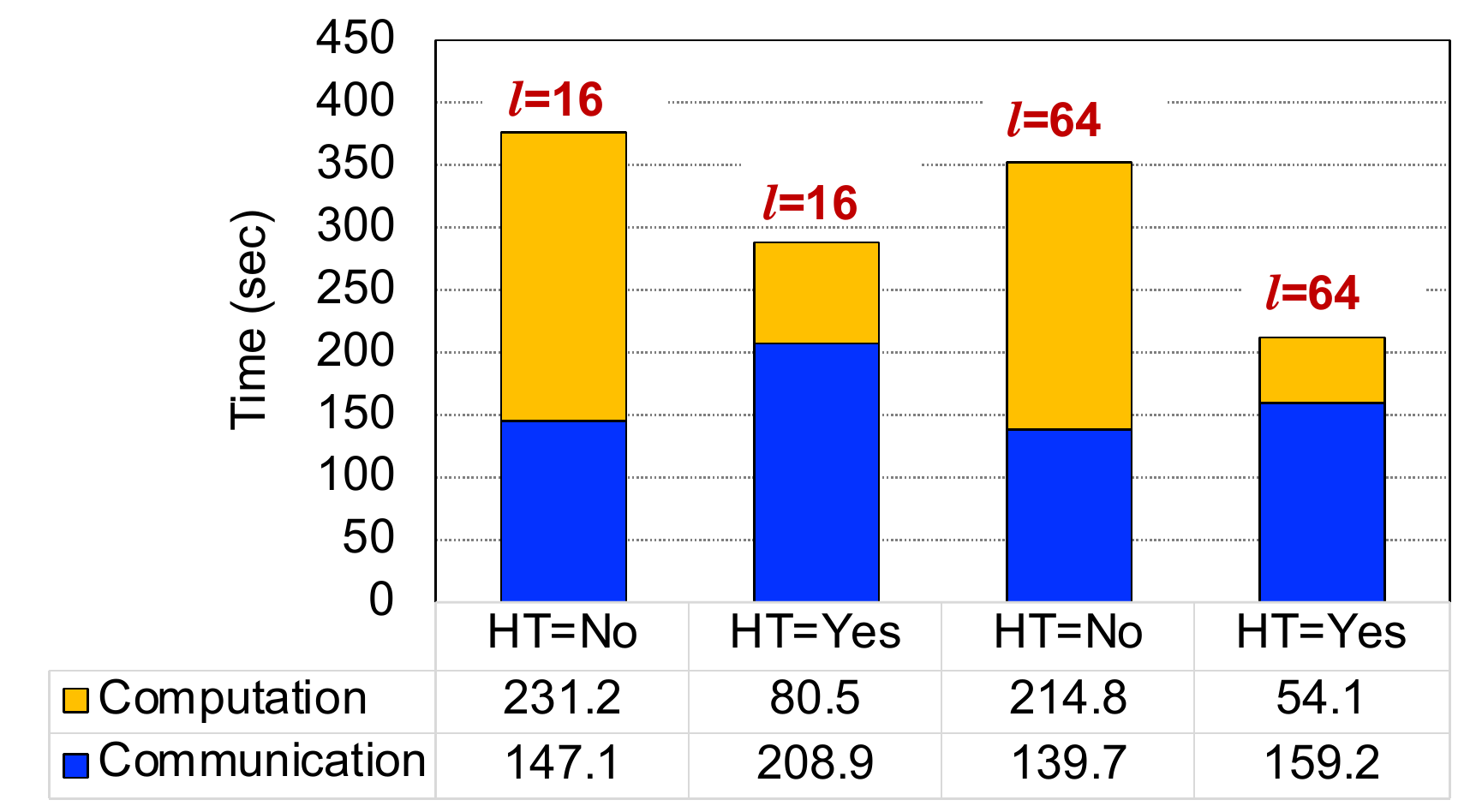} 
    \caption{The impact of using hyperthreads when squaring Metaclust50 on 4096 nodes on Cori-KNL. HT=Yes means 4 hardware threads per core are used. At this setting with 4096 nodes, we use 256 threads per node totaling 1,048,576 threads.  HT=No means one thread per core is used, totaling 262,144 threads. As before, we used 16 threads per process. The number of layers used in the 3D grid is shown at the top of each bar. For this matrix, hyperthreading reduces the computation time, but increases communication time. Overall, hyperthreading decreases the total runtime of this SpGEMM.
    }
    \vspace{-5pt}
    \label{fig:hypethread}
\end{figure}

\begin{figure}[!t]
    \centering
    \includegraphics[width=.75\linewidth]{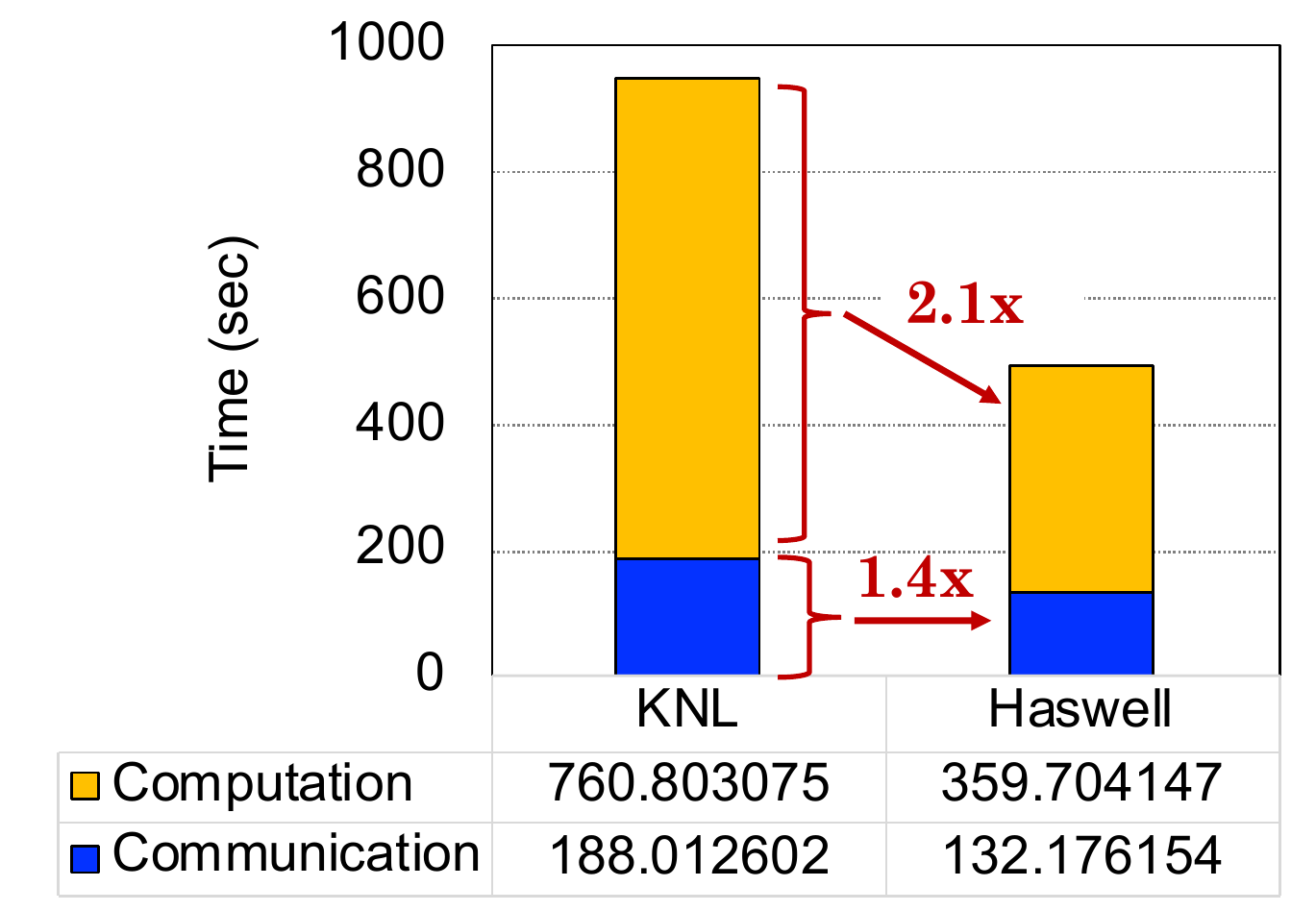} 
    \caption{Squaring Isolates-small on 256 nodes of Cori-KNL and Cori-Haswell. While the communication network remains the same, Cori-Haswell uses 32 fast cores per node. We use 6 cores/process on Haswell and 16  cores/process on KNL. With 16 layers and 23 batches, both experiments use the same process grid. Arrowheads show that computation and communication becomes $2.1\times$ and $1.4\times$ faster on Haswell. 
    }
    \vspace{-5pt}
    \label{fig:haswell}
\end{figure}

\subsection{Impact of hyper-threading at extreme scale}
\vspace{-2pt}
As with most modern manycore processors, each KNL processor has four hardware threads. 
In our prior experiments, we did not use hyperthreading because it can increase the communication time significantly due to larger process grids. 
Here, we study the impact of hyperthreads when squaring Metaclust50 on 4096 nodes on Cori-KNL.
Fig.~\ref{fig:hypethread} shows that hyperthreading can help SpGEMM run faster by reducing the computation time significantly even though the communication time may increase. 
This gives us an unprecedented scalability to more than one million threads, which will help scale many graph and machine learning applications to upcoming exascale systems.  
The impact of hyperthreading is more significant when the total runtime is dominated by computation (for example when $l=64$ in Fig.~\ref{fig:hypethread}).
That means, hyperthreading may not help when SpGEMM becomes communication-bound.

\subsection{Impact of using faster processors}
If we employ faster processors while using the same communication network, communication could quickly become the bottleneck.
Hence, faster in-node computations are expected to make \textproc{BatchedSumma3D} more even more beneficial. 
We investigate this hypothesis by running \textproc{BatchedSumma3D} on 256 nodes of Cori-KNL and Cori-Haswell and show the result in Fig.~\ref{fig:haswell}.
We observe that computation is about $2.1\times$ faster on Haswell.
Even with the same communication network, communication on Cori-Haswell is $1.4\times$ faster possibly because of faster data possessing around MPI calls. 
Since communication does not scale as well as computations on Haswell,  communication takes an increased fraction of the total time in comparison to KNL.
We expect even better benefits of our algorithm on GPU-based clusters because of the availability of faster in-node computations.

\begin{figure}[!t]
    \centering
    \includegraphics[width=.66\linewidth]{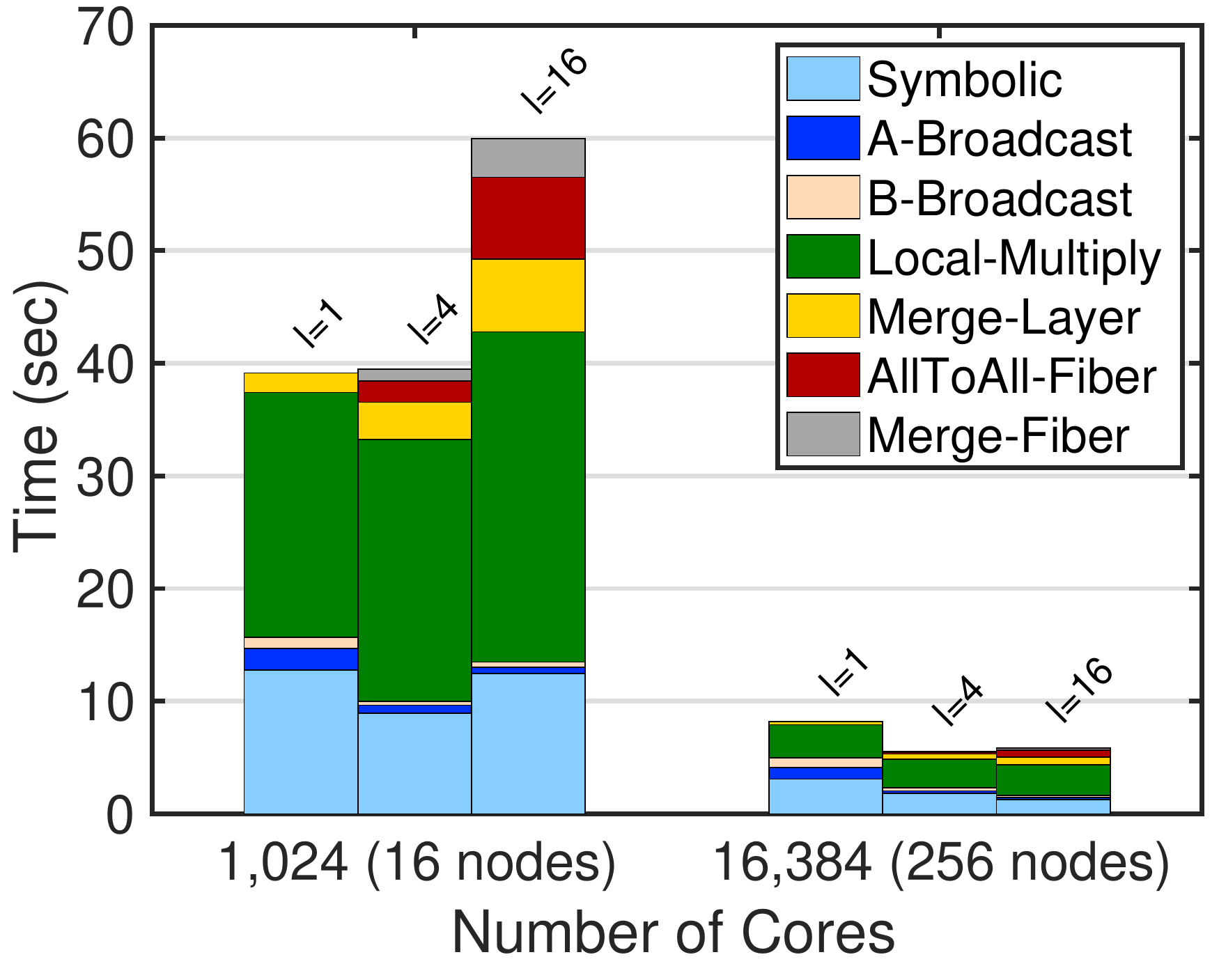} 
    \vspace{-5pt}
    \caption{Squaring Eukarya, the smallest matrix in our test suite, on Cori-KNL. SpGEMM needs two batches when $l=16$ on 16 nodes. For all other cases, $b=1$. \textproc{SUMMA3D} is not useful when communication cost is insignificant on 16 nodes. However, \textproc{SUMMA3D} with 4 layers is still useful on 256 nodes even though batching is not needed for this small matrix.
    }
    \vspace{-10pt}
    \label{fig:eukarya}
\end{figure}

\subsection{Applicability with small matrices at low concurrency}
As observed in previous experiments, the \textproc{BatchedSumma3D} algorithm is extremely effective when multiplying large-scale matrices on thousands of nodes.
Fig.~\ref{fig:eukarya} demonstrates that \textproc{BatchedSumma3D} can reduce the A-Broadcast time even on 16 nodes.
However, the reduced communication time has little impact on the total runtime when communication does not dominate the runtime. 
On 256 node, \textproc{BatchedSumma3D} runs faster with 4 layers. 
However, using 16 layers on 256 nodes does not reduce the runtime any further as AllToAll-Fiber starts to become the communication bottleneck.
Hence, \textproc{BatchedSumma3D} is useful even on few hundred nodes if $l$ is set to a small value.  

\begin{figure}[!t]
    \centering
    \includegraphics[width=.95\linewidth]{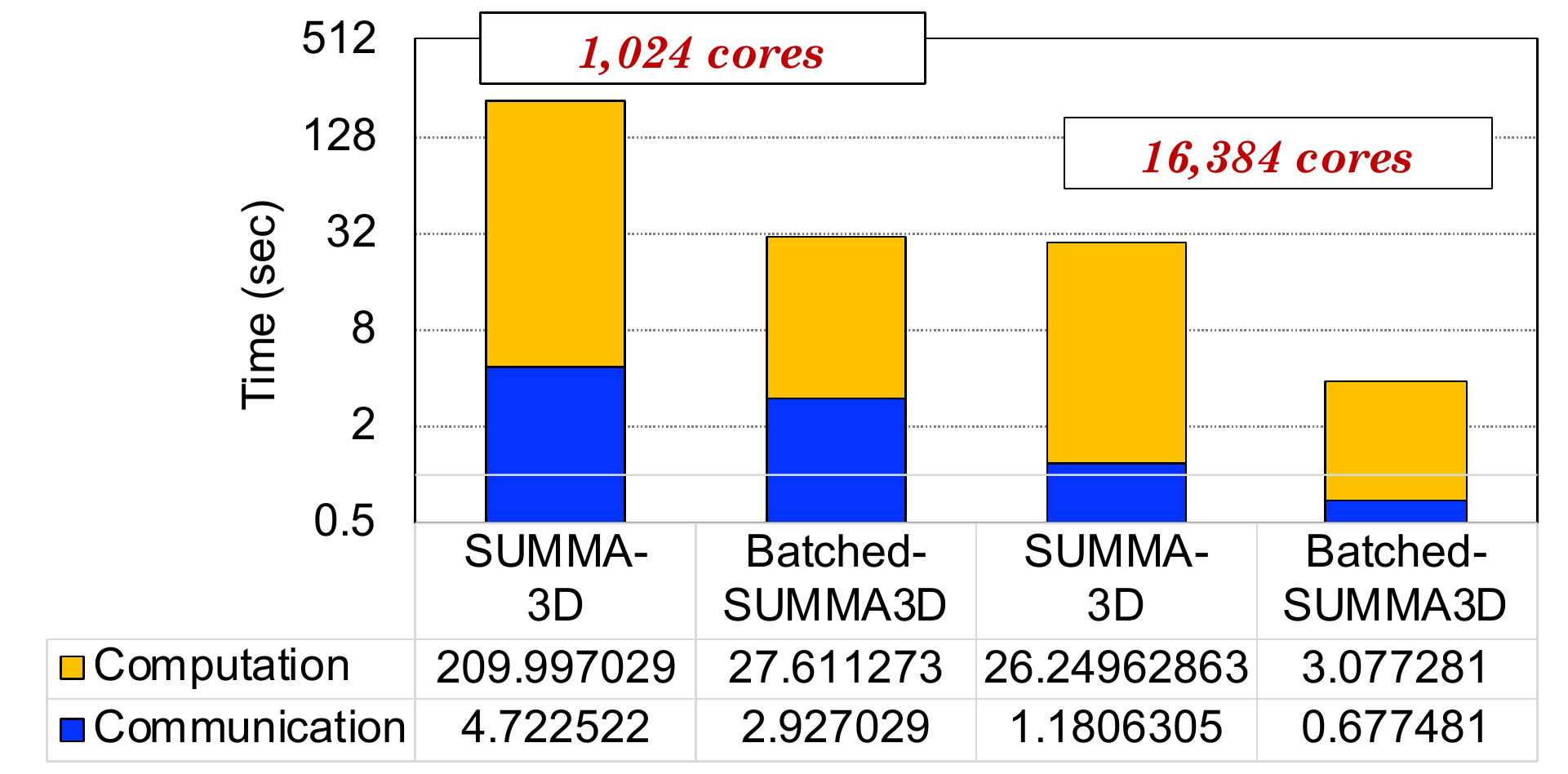} 
    \caption{\revision{Comparing  \textproc{BatchedSumma3D} developed in this paper with the SUMMA3D algorithm presented in \cite{Azad2016} in squaring the Eukarya matrix with 4 layers without batching. 
    }
    }
    \vspace{-10pt}
    \label{fig:compare}
\end{figure}

\subsection{A direct comparison with 3D Sparse SUMMA}
Fig.~\ref{fig:compare} compares 
\textproc{BatchedSumma3D} with SUMMA3D presented in previous work~\cite{Azad2016}. We downloaded the SUMMA3D code from the CombBLAS library.
Note that the previous SUMMA3D algorithm simply fails when the memory requirement exceeds avaiable memory.
Here, we square the Eukarya matrix where batching is not needed on 16 nodes and 256 nodes of Cori KNL using 4 layers and 16 threads per process. The computation in \textproc{BatchedSumma3D} is more than 8$\times$ faster than the previous work, while the communication is also slightly faster. 
We relied on new hash-based multiplication and merging algorithms (see Sec~\ref{sec:summa3dbatch}), which made the computation much faster.

\section{Conclusions}
This paper presents a robust SpGEMM algorithm that can multiply matrices in batches even when the output matrix exceeds the available memory of large supercomputers. 
Additionally, the presented algorithm reduces the communication so that SpGEMM can scale to the limit of modern supercomputers. 
These two techniques together eliminate two fundamental barriers -- memory and communication -- in large-scale sparse data analysis. 

Our result is unexpectedly positive because communication-avoiding (CA) matrix multiplication algorithms trade increased memory for reduced communication. The conventional wisdom suggests that CA algorithms would be detrimental in this already memory-constrained regime. However, 3D algorithms offset the increased broadcast costs associated with batching required in the memory constrained setting, creating a previously unexplored synergy.

Our algorithm will boost many applications in genomics, scientific computing, and social network analysis where SpGEMM has emerged as a key computational kernel.
For example, Yelick et al.~\cite{yelick2020parallelism} regarded SpGEMM as a parallelism motif of genomic data analysis with applications in alignment, profiling, clustering and assembly for both single genomes and metagenomes.
With the size of genomic data growing exponentially, extreme-scale SpGEMM presented in this paper will enable rapid scientific discoveries in these applications. 

\section*{Acknowledgments}
This work is supported in part by the Advanced Scientific Computing Research (ASCR) Program of the Department of Energy Office of Science under contract
No. DE-AC02-05CH11231, and in part by the Exascale Computing Project
(17-SC-20-SC), a collaborative effort of the U.S. Department
of Energy Office of Science and the National Nuclear Security
Administration.

This work used resources of the NERSC supported by the Office of Science of the
DOE under Contract No. DEAC02-05CH11231.

\bibliographystyle{IEEEtran}
\bibliography{batched3D}
\end{document}